\documentclass[12pt,a4paper]{article}
\usepackage{amsmath,amssymb,amsthm}
\usepackage[margin=1.0in]{geometry}
\usepackage{cite}
\usepackage{graphicx}
\usepackage{enumerate}
\usepackage{epsfig}
\usepackage{textcomp}
\usepackage{float}
\usepackage{bm}
\usepackage[displaymath, mathlines]{lineno}
\allowdisplaybreaks[0]
\floatstyle{plaintop}
\restylefloat{table}
\usepackage[colorlinks=true
,urlcolor=blue
,anchorcolor=blue
,citecolor=blue
,filecolor=blue
,linkcolor=blue
,menucolor=blue
,linktocpage=true
,pdfproducer=medialab
,pdfa=true
]{hyperref}
\numberwithin{equation}{section}

\usepackage{adjustbox}
\usepackage{lipsum}
\usepackage{pbox}

\usepackage{graphicx}
\usepackage{subcaption}

\usepackage[font=footnotesize]{caption}

\def\be{\begin{equation}}
\def\ee{\end{equation}}
\def\bea{\begin{eqnarray}}
\def\eea{\end{eqnarray}}

\def\pa{\partial}

\def\lp{\left(}
\def\rp{\right)}
\def\ls{\left[}
\def\rs{\right]}
\def\nn{\nonumber}
\def\ie{{\it i.e., }}

\makeatletter
\renewcommand\section{\@startsection {section}{1}{\z@}%
	{-3.5ex \@plus -1ex \@minus -.2ex}
	{2.3ex \@plus.2ex}%
	{\normalfont\large\bfseries}}
\renewcommand\subsection{\@startsection{subsection}{2}{\z@}%
	{-3.25ex\@plus -1ex \@minus -.2ex}%
	{1.5ex \@plus .2ex}%
	{\normalfont\bfseries}}
\makeatother

\begin{document}

\begin{center}
\addtolength{\baselineskip}{.5mm}
\thispagestyle{empty}
\begin{flushright}
\end{flushright}

\vspace{20mm}

{\Large \bf Novel thermodynamic inequality for rotating AdS black holes}
\\[5mm]
{Hamid R. Bakhtiarizadeh\footnote{h.bakhtiarizadeh@kgut.ac.ir}}
\\[5mm]
{\it Department of Nanotechnology, Graduate University of Advanced Technology, Kerman, Iran}

\vspace{20mm}

{\bf  Abstract}
\end{center}

	We propose a new thermodynamic inequality for stationary, asymptotically Anti-de Sitter rotating black holes, expressed in terms of their angular momentum, mass and thermodynamic volume as $ 4\pi J^2/(3MV)<1 $. For the solutions considered, this inequality follows from the requirement that the thermodynamic identities admit a physical horizon with positive area. We examine Kerr(-Newman)-AdS, Kerr-Sen-AdS, pairwise-equal four-charge AdS black holes, charged rotating AdS black strings, and charged rotating AdS C-metrics, finding strong supporting evidence for the inequality across different horizon topologies. We then reformulate the inequality in terms of the underlying metric parameters and compare the resulting constraints with the corresponding regularity and horizon-existence conditions. This analysis shows that the proposed inequality has different physical content for different geometries. We further examine charged super-entropic black holes, for which the ultraspinning limit leads instead to a lower bound on the thermodynamic ratio. This result is consistent with the violation of the reverse isoperimetric inequality characteristic of super-entropic solutions. Moreover, when combined with the refined reverse isoperimetric inequality, the proposed bound ensures the physicality of the Kerr-AdS reference solution entering the refined bound and thereby yields the conventional reverse isoperimetric inequality. We also extend the proposed relation to higher dimensions and test it for singly spinning thin AdS black ring with nonspherical horizon topology, finding that the corresponding inequalities are satisfied throughout the thin-ring regime.
	
\vfill
\newpage


\section{Introduction}\label{int}

The thermodynamics of black holes provides a remarkable framework for exploring the interplay between gravitation, quantum theory, and thermodynamics. The identification of black hole entropy with the area of its event horizon by Bekenstein \cite{Bekenstein:1973ur}, together with Hawking's discovery of black-hole radiation \cite{Hawking:1975vcx}, established a fundamental connection between gravitational dynamics and thermodynamics. In asymptotically Anti-de Sitter (AdS) spacetimes, these ideas acquire further significance through the AdS/CFT correspondence \cite{Maldacena:1997re}, which provides a framework for relating gravitational systems to properties of a dual conformal field theory and has stimulated further investigations of entropy bounds in gravitational systems.

A major extension of black hole thermodynamics is obtained by promoting the cosmological constant to a thermodynamic pressure and introducing its conjugate thermodynamic volume \cite{Kastor:2009wy,Cvetic:2010jb}. This extended phase space formulation has led to the development of ``black hole chemistry'' \cite{Karch:2015rpa} and has revealed a rich thermodynamic structure, including critical behavior and phase transitions \cite{Kubiznak:2012wp}, as well as thermodynamic cycles in which black holes can act as heat engines \cite{Johnson:2014yja}. The introduction of thermodynamic volume has also motivated studies of its microscopic and holographic interpretation and its relation to the degrees of freedom of the dual theory \cite{Johnson:2019wcq}. These developments have broadened the scope of black hole thermodynamics and have led to investigations of holographic complexity \cite{AlBalushi:2020rqe}, chemical potentials in holographic thermodynamics \cite{Visser:2021eqk}, and the higher-dimensional structure of black hole thermodynamics \cite{Frassino:2022zaz,Ahmed:2023snm}. At the same time, extended thermodynamics has been explored in lower-dimensional settings \cite{Frassino:2015oca}, as well as in theories with higher-curvature corrections and Chern-Simons terms \cite{Frassino:2019fgr}. For a comprehensive review of black hole chemistry, see Ref.~\cite{Mann:2025xrb}.

The extended thermodynamic framework is particularly useful for studying black holes with nontrivial rotational and horizon structures. Investigations of rotating black holes and black rings have revealed rich phase behavior and nontrivial thermodynamic volumes associated with more complicated horizon geometries \cite{Altamirano:2014tva}. Related developments have considered de Sitter backgrounds \cite{Dolan:2013ft}, accelerating black holes \cite{Gregory:2019dtq}, and black holes with unusual horizon topologies \cite{Klemm:2014rda}. The structure of gauged-supergravity solutions \cite{Gnecchi:2013mja} provides further examples in which thermodynamic quantities and geometric properties are intertwined. These results emphasize the importance of understanding thermodynamic and geometric inequalities across a broad range of black hole solutions rather than restricting attention to the simplest spherical-horizon geometries.

One of the central geometric constraints arising in this context is the reverse isoperimetric inequality (RII), which conjectures that, for a fixed thermodynamic volume, the entropy of a black hole is bounded from above by that of the Schwarzschild-AdS black hole \cite{Dolan:2013ft,Feng:2017wvc}. Equivalently, the corresponding dimensionless isoperimetric ratio is bounded from below by unity. The RII has been verified for a wide range of conventional black hole solutions and has provided a useful criterion for relating black hole entropy to thermodynamic volume. Its status becomes more subtle for rotating solutions with ultraspinning behavior and for black holes with nonstandard horizon geometries. In particular, super-entropic black holes obtained through an ultraspinning limit violate the conventional RII \cite{Hennigar:2014cfa,Johnson:2019mdp}. Such solutions provide an important setting in which to investigate the conditions underlying entropy bounds and to understand how thermodynamic inequalities depend on the geometry and topology of the horizon. The stability and physical properties of these exotic solutions have also been investigated in the literature \cite{Johnson:2019mdp,Appels:2019vow}.

A refined form of the reverse isoperimetric inequality for rotating black holes was proposed in Ref.~\cite{Amo:2023bbo}. This refined relation introduces an area bound involving the mass, angular momentum, and thermodynamic volume and compares a general rotating black hole with an appropriate Kerr-AdS reference solution. Such a refinement is particularly relevant for rotating black holes, for which the usual comparison with the Schwarzschild-AdS geometry does not explicitly account for angular momentum. More generally, refined versions of geometric entropy bounds have also been investigated for lower-dimensional quantum black holes; in particular, refined versions of the quantum Penrose inequality and the reverse isoperimetric inequality were proposed for known three-dimensional asymptotically AdS quantum black holes in Ref.~\cite{Frassino:2024bjg}. These developments motivate a closer examination of thermodynamic combinations involving mass, angular momentum, and thermodynamic volume and of their relation to geometric inequalities.

In this work, we propose and investigate a thermodynamic inequality for stationary, asymptotically AdS rotating black holes of the form
\begin{equation}
	\frac{4\pi J^2}{3MV}<1.
\end{equation}
We examine this relation for a broad class of four-dimensional rotating AdS solutions, including Kerr-AdS, Kerr-Newman-AdS (KN-AdS), Kerr-Sen-AdS (KS-AdS), pairwise-equal-charge rotating black holes in gauged supergravity, charged rotating AdS black strings, and charged accelerating AdS black holes. For the solutions considered, the inequality follows from the requirement that the relevant thermodynamic relations admit a physical configuration with positive horizon area. We subsequently reformulate the inequality in terms of the underlying metric parameters and compare the resulting constraints with the independent conditions associated with rotational regularity and the existence of an event horizon. This analysis allows us to distinguish cases in which the proposed inequality reproduces an existing regularity condition from cases in which it provides an additional restriction on the physical parameter space.

The proposed inequality also has a direct connection with the refined reverse isoperimetric inequality. When combined with the refined bound of Ref.~\cite{Amo:2023bbo}, the proposed thermodynamic inequality guarantees that the Kerr-AdS reference solution entering that bound is physical and thereby leads to the conventional RII. We emphasize that this conclusion is conditional on the refined inequality of Ref.~\cite{Amo:2023bbo}; the proposed thermodynamic inequality alone does not constitute an independent derivation of either the refined bound or the conventional RII. This distinction is important for understanding the precise role played by the proposed inequality in the thermodynamic description of rotating AdS black holes.

We further examine charged super-entropic black holes, which provide a particularly relevant test because they violate the conventional RII. In this case, the ultraspinning limit leads to a qualitatively different behavior of the proposed thermodynamic ratio. Rather than yielding the upper bound characteristic of the regular compact-horizon solutions, positivity of the horizon area leads to a lower bound on the ratio. This condition is consistent with the known violation of the reverse isoperimetric inequality by super-entropic solutions and demonstrates that the physical interpretation of the proposed thermodynamic relation depends on the underlying geometry and horizon topology.

Finally, we investigate a higher-dimensional extension of the proposed relation. Since this extension is motivated by higher-dimensional thermodynamic relations together with refined reverse isoperimetric bounds, we present it as a conjectural extension rather than as a general theorem. As a nontrivial test beyond spherical horizon geometries, we consider thin asymptotically AdS black rings with nonspherical horizon topology. The corresponding inequalities are satisfied throughout the thin-ring regime in both even and odd spacetime dimensions. These results provide evidence for the applicability of the proposed thermodynamic relation beyond the four-dimensional examples considered above, while also demonstrating that its detailed form and physical interpretation can depend on the spacetime dimension and horizon topology.

The paper is organized as follows. In Sec.~\ref{evid}, we review the evidence for the proposed rotational thermodynamic inequality and derive it for several rotating AdS black-hole geometries, including Kerr-AdS, uncharged KS-AdS, and uncharged rotating black strings. In Sec.~\ref{reform}, we reformulate the inequality in terms of the underlying metric parameters and distinguish the resulting regularity constraints from the independent conditions required for the existence of an event horizon. We then consider further four-dimensional tests in Sec.~\ref{fur}, including KN-AdS, charged KS-AdS, charged rotating black strings, charged accelerating AdS black holes, and pairwise-equal-charge rotating black holes in gauged supergravity. The super-entropic case is studied separately in Sec.~\ref{su}, where we examine the effect of the ultraspinning limit and its relation to the reverse isoperimetric inequality. In Sec.~\ref{rev}, we discuss the connection between the proposed thermodynamic inequality and the refined reverse isoperimetric inequality, clarifying the assumptions entering this connection. Finally, in Sec.~\ref{ext}, we discuss the higher-dimensional extension and test the proposed inequality for thin asymptotically AdS black rings with nonspherical horizon topology. We conclude in Sec.~\ref{conc} with a summary of our main results and their implications.

\section{Evidence for the inequality}\label{evid}

We first consider the Kerr-AdS black hole \cite{Carter:1968ks,Hawking:1998kw,Gibbons:2004uw}, for which the thermodynamics in the extended phase space was first studied in \cite{Cvetic:2010jb}. In this framework, the following identity among the thermodynamic variables holds \cite{Amo:2023bbo}:
\be\label{K-AdSidentity}
36 \pi M^2 V^2 -M^2 A^3 -64 \pi^3 J^4=0.
\ee
It follows directly from the corresponding KN-AdS identity (\ref{KN-AdSiden}) upon setting $ Q=0 $. Here, \(A\) is the horizon area, \(V\) is the thermodynamic volume, \(M\) is the mass (enthalpy), and \(J\) is the angular momentum of the black hole. The identity can be solved explicitly for the horizon area, yielding
\be
A^3=36\pi V^2\ls1-\lp\frac{4\pi J^2}{3MV}\rp^2\rs.
\ee
Requiring a positive horizon area, $ A^3>0 $, leads to the following \emph{rotational thermodynamic inequality} (RTI)
\be\label{KAineq}
{\cal I}\equiv\frac{4\pi J^2}{3MV}<1.
\ee

We next consider the KS-AdS black hole. Sen originally discovered a charged, rotating black hole solution in the low-energy limit of heterotic string theory, now commonly referred to as the Kerr-Sen black hole \cite{Sen:1992ua}. Its thermodynamics in the extended phase space was subsequently studied in \cite{Wu:2020cgf,Sharif:2021yis}. Starting from the standard first law of black hole thermodynamics in the extended phase space, given in the first line of Eq. (\ref{stfs}), we obtain the thermodynamic volume of the KS-AdS black hole in terms of the thermodynamic potentials as
\bea\label{KSAthermovol}
{\widetilde V}=\lp\frac{\pa \widetilde M}{\pa P}\rp_{S,J,Q}=\frac{4 \sqrt{S} \left\{6 \pi ^2 J^2+S \left[S \left(8 P S+3\right)+3 \pi  Q^2\right]\right\}}{3 \sqrt{\pi }\sqrt{8 P S+3} \sqrt{12 \pi ^2 J^2+S \left[S \left(8 P S+3\right)+6 \pi  Q^2\right]}}.
\eea
Solving Eq. (\ref{KSAthermovol}) for the pressure yields
\bea
&& P= \frac{1}{8 S^3 \left(16 S^3-9 \pi  {\widetilde V}^2\right)}\left[3 \left(3 \pi ^{3/2} {\widetilde V} \sqrt{9 \pi {\widetilde V}^2-16 S^3} \left(2 \pi  J^2+Q^2 S\right)\right.\right.\nn\\&&\left.\left.+2 \pi ^2 J^2 \left(9 \pi {\widetilde V}^2-16 S^3\right)-16 \pi  Q^2 S^4+9 \pi ^2 Q^2 S {\widetilde V}^2-16 S^5+9 \pi  S^2 {\widetilde V}^2\right)\right].
\eea
Substituting this expression into the mass formula (\ref{KSAM}) and using the Bekenstein-Hawking relation, $ S=A/4 $, we eliminate the pressure and reexpress the mass $ {\widetilde M} $ in terms of $ (A, {\widetilde V}, J, Q) $, yielding the following identity:
\be\label{KS-AdSiden}
36 \pi {\widetilde M}^2 {\widetilde V}^2 -{\widetilde M}^2 A^3 -64 \pi^3 J^4=16 \pi^2 Q^2 J^2 A.
\ee
By setting $ Q=0 $ we find the corresponding identity for uncharged KS-AdS black hole as 
\be\label{UKS-AdSiden}
36 \pi {\widetilde M}^2 {\widetilde V}^2 -{\widetilde M}^2 A^3 -64 \pi^3 J^4=0.
\ee	
It can be seen from Eq. (\ref{KS-AdSiden}) that the identity satisfied by the thermodynamic variables has the same structure as that for the KN-AdS black hole, with the mass \(M\) and thermodynamic volume \(V\) replaced by \(\widetilde{M}\) and \(\widetilde{V}\), respectively, as given in Eq. \eqref{restmassangthermovol}. This correspondence arises because both the standard first law of black-hole thermodynamics and the Bekenstein-Smarr mass formula must be satisfied simultaneously. Consequently, the RTI takes the following form:
\be\label{UKSAineq}
{\cal I}\equiv\frac{4\pi J^2}{3{\widetilde M}{\widetilde V}}<1,
\ee

A similar algebraic structure arises in the extended thermodynamics of asymptotically AdS rotating black string solutions introduced by Lemos \cite{Lemos:1994xp,Lemos:1995cm}. For the charged solutions, however, obtaining an explicit identity among the thermodynamic variables $ (\cal M, \cal A, \cal V, \cal J, \cal Q) $ per unit length of the black string horizon requires solving a polynomial equation of high degree. For the uncharged solutions, on the other hand, we can reexpress the mass of the black string in terms of the thermodynamic variables, $ {\cal S} $, $ P $ and $ {\cal J} $ as \cite{Bakhtiarizadeh:2025uks}:
\be\label{bsmass}
\mathcal{M}=\frac{4 \sqrt{P} \left(\pi ^3 \mathcal{J}^4+\pi ^{3/2} \mathcal{J}^2 \sqrt{\pi ^3 \mathcal{J}^4+54 P \mathcal{S}^6}+18 P \mathcal{S}^6\right)}{\sqrt{3} \sqrt[4]{\pi } \left(\pi ^{3/2} \mathcal{J}^2+\sqrt{\pi ^3 \mathcal{J}^4+54 P \mathcal{S}^6}\right)^{3/2}},
\ee
which can be obtained from Eq.~\eqref{mass} by setting $ {\cal Q}=0 $. From the first law of black hole thermodynamics, the thermodynamic volume per unit length of the black string horizon is readily obtained as
\be\label{thermovol}
\mathcal{V}=\frac{\sqrt{\pi ^{3/2} \mathcal{J}^2+\sqrt{\pi ^3 \mathcal{J}^4+54 P \mathcal{S}^6}}}{\sqrt{3} \sqrt[4]{\pi } \sqrt{P}}.
\ee
Solving the above equation for the pressure yields 
\be\label{unchpressure}
P=\frac{2 \pi \mathcal{J}^2}{3  \mathcal{V}^2}+\frac{6 \mathcal{S}^6}{\pi \mathcal{V}^4}.
\ee
Substituting the above expression into the mass formula (\ref{bsmass}), we eliminate the pressure and reexpress the mass in terms of $ (\cal S, \cal V, \cal J) $
\be\label{unchmass}
\mathcal{M}= \frac{4 \pi  \mathcal{J}^2}{3 \mathcal{V}}+\frac{8 \mathcal{S}^6}{\pi  \mathcal{V}^3}.
\ee
Using the Bekenstein-Hawking relation, it is straightforward to show that, in terms of the thermodynamic variables \((\mathcal{M},\mathcal{A},\mathcal{V},\mathcal{J})\), the following identity holds for asymptotically AdS uncharged rotating black strings:
\be\label{BSiden}
3{\cal A}^6-512\pi {\cal V}^2\left(3 {\cal M}{\cal V}-4\pi {\cal J}^2\right)=0.
\ee
This can be written as
\be
{\cal A}^6=\frac{512\pi {\cal V}^2}{3}\left(3 {\cal M}{\cal V}-4\pi {\cal J}^2\right).
\ee
Thus, requiring a positive horizon area, \(\mathcal{A}>0\), leads to
\be\label{Urbsineq}
{\cal I}\equiv\frac{4\pi {\cal J}^2}{3{\cal M}{\cal V}}<1.
\ee
This provides further supporting evidence for the proposed RTI in a solution with a non-spherical horizon topology. The similarity of the algebraic structures observed for the solutions considered suggests that the same thermodynamic combination may arise across different horizon topologies.

\section{Reformulation in terms of metric parameters}\label{reform}

To establish that the proposed inequality is not merely equivalent to horizon regularity within a given family, but instead may encode an independent physical constraint, it is useful to determine whether it contains information beyond that already implied by the horizon quantities. To this end, we reformulate the RTI directly in terms of the primary parameters of the metric and compare the resulting constraint with the condition for horizon regularity. We illustrate this analysis for three uncharged rotating AdS geometries with distinct horizon structures given in the previous section.

Substituting the Kerr-AdS thermodynamic quantities into the inequality, which can be obtained from the corresponding KN-AdS quantities given in Appendix~\ref{KNblackhole} by setting $ q=0 $, yields
\be
	{\cal I}
	=
	\frac{a^2(\ell^2+r_+^2)}
	{2\ell^2r_+^2+a^2(\ell^2-r_+^2)}<1.
\ee
Consequently, for a positive horizon radius, $ r_+>0 $, the inequality \eqref{KAineq} implies that
\be\label{genonsing}
\lvert a \rvert<\ell.
\ee
Since $ J=aM $, this can equivalently be written as
\be\label{genonsing1}
|J|<M\ell,
\ee
or equivalently $ \Xi>0 $. Thus, in the Kerr-AdS family, the proposed inequality is not an additional independent restriction on the black-hole parameter space; rather, it provides a thermodynamic representation of the standard regularity condition associated with the Kerr-AdS geometry.

It is important to distinguish this condition from the condition for the existence of a positive horizon. The location of the horizon $ r_+ $ is determined by the largest root of the horizon equation $ \Delta_{r}(r_+)=0 $. To ensure the existence of horizons, the mass parameter must satisfy
\be
m\geq m_{\rm ext}(a,\ell),
\ee
where
\begin{equation}
		m_{\rm ext}
	=
	\frac{(r_{\rm ext}^2+a^2)
		(\ell^2+r_{\rm ext}^2)}
	{2r_{\rm ext}\ell^2}
	,
	\qquad
r_{\rm ext}^2
=
\frac{
	\sqrt{\lp a^2+\ell^2\rp^2+12 a^2\ell^2}
	-\lp a^2+\ell^2\rp
}{6}.
\end{equation}

For the uncharged KS-AdS black hole, using the thermodynamic quantities given in Appendix \ref{KSAblackhole} with $ b=q=0 $, together with the horizon equation $ \Delta_r(r_+)=0 $, the RTI \eqref{UKSAineq} takes the following form
\be
{\cal I}=
\frac{
	a^2\ell^2\left(\ell^2+r_+^2\right)
}{
	(a^2+\ell^2)
	\left[
	a^2(\ell^2-r_+^2)
	+2\ell^2r_+^2
	\right]
}<1,
\ee
which implies that
\be
\frac{2(\ell^2-a^2)(\ell^2+a^2)}{\ell^2}>0.
\ee
Therefore, the inequality implies that $ \lvert a \rvert<\ell $.

It is important to distinguish between the mass \(M\) and the thermodynamic quantity \(\widetilde{M}\) defined in Eq. (\ref{restmassangthermovol}). In the uncharged limit, \(b=q=0\), the KS-AdS solution reduces to the Kerr-AdS black hole, for which $ J=aM $. Consequently, the standard Kerr-AdS condition \(|a|<\ell\) can be expressed in terms of the mass \(M\) as $ |J|<M\ell $, or equivalently $ \Xi>0 $. On the other hand, using $ \widetilde{M}=M+aJ/\ell^2 =M\left(1+a^2/\ell^2\right) $, the same condition, on the physical branch \(|a|<\ell\), can equivalently be written as
\be
|J|<\frac{1}{2}\widetilde{M}\ell.
\ee
We therefore retain \(\widetilde{M}\) in the definition of the rotational thermodynamic inequality, since the corresponding identity and the inequality for the KS-AdS solution are naturally expressed in terms of \((\widetilde{M},\widetilde{V},J)\). It is worth noting that, to the best of our knowledge, the above bound has not previously been reported in the literature for KS-AdS black holes, where $ \widetilde{M} $ denotes the thermodynamic mass entering the first law. This provides a simple bound on the angular momentum in terms of the physical mass and the AdS length scale.

The condition for the existence of an event horizon can be expressed as
\be
m\geq m_{\rm ext}(a,q,\ell)
\ee
with the corresponding extremal mass given by
\be
m_{\rm ext}
=
\frac{
	(r_{\rm ext}^2+a^2)
	\left(\ell^2+r_{\rm ext}^2\right)
}{
	2\ell^2 r_{\rm ext}
},
\ee
and $ r_{\rm ext} $ is the positive real root of the extremality equation
\be
3r_{\rm ext}^{4}
+\left(a^2+\ell^2\right)r_{\rm ext}^{2}
-a^2\ell^2=0.
\ee

Thus, for the uncharged KS-AdS solution, the proposed inequality is satisfied on the standard regular branch and reduces to the simple bound $ |J|<\widetilde{M}\ell/2 $, which, to the best of our knowledge, has not previously been reported in the literature for this solution.

A complementary example is provided by the uncharged rotating black string. In this case, the RTI \eqref{Urbsineq} reduces to
\be
{\cal I}
	=
	\frac{3a^2}{3a^2+2\ell^2}<1.
\ee

For an asymptotically AdS solution, $ \ell^2>0 $, and therefore, for any finite $ a $, the denominator is strictly greater than the numerator and consequently the inequality is automatically satisfied throughout the physical parameter space of the uncharged rotating black string. It is also instructive to express the latter result entirely in terms of physical conserved quantities. Eliminating the rotation parameter $ a $, one obtains
\be
{\cal I}
=
\frac{3-\sqrt{
	9-\dfrac{8{\cal J}^2}{{\cal M}^2\ell^2}
}
}{2}<1,
\ee
which is precisely the bound $ |{\cal J}|<{\cal M}\ell $, given in Eq. \eqref{genonsing1}. To the best of our knowledge, this provides the first explicit derivation of the bound as a thermodynamic regularity condition for the uncharged rotating black string.

The horizon equation gives
\be
m=\frac{r_+^3}{\ell^2},
\ee
so that
\be
r_+=(m\ell^2)^{1/3}.
\ee
Thus, for $ m>0 $, a positive horizon exists without a Kerr-like extremality threshold. 

These examples demonstrate that the RTI can be recast as a bound involving the physical conserved quantities, while its precise physical content depends on the geometry under consideration. For Kerr-AdS, the inequality reduces to the standard regularity condition $ |a|<\ell $, and is equivalently expressed as $ |J|<M\ell $. For KS-AdS, on the other hand, the inequality yields the bound $ |J|<\widetilde{M}\ell/2 $, where $\widetilde{M}$ is the thermodynamic mass entering the first law. Finally, for the uncharged rotating black string, the inequality gives $ |{\cal J}|<{\cal M}\ell $, which is automatically satisfied throughout the physical parameter space for finite rotation parameter $a$. Thus, while the RTI reduces to a familiar regularity condition for Kerr-AdS and yields a nontrivial mass-angular-momentum bound for KS-AdS and uncharged rotating AdS black string, it is automatically satisfied for the uncharged rotating black string and therefore does not impose an additional restriction on its rotation parameter.

\section{Further tests of the inequality}\label{fur}

The treatment of the KN-AdS black hole in terms of the cubic polynomial
\be
f(A)\equiv M^2 A^3+16\pi^2 Q^2J^2A-36\pi M^2V^2+64\pi^3J^4,
\ee
is well defined. Since
\be
f'(A)=3M^2A^2+16\pi^2Q^2J^2>0,
\ee
for physically meaningful parameters ($M\neq0$ and real $Q,J$), $f(A)$ is strictly increasing and therefore has exactly one real root. This root is positive if and only if $f(0)<0$, which is precisely the RTI \eqref{KAineq}.

On the other hand, expressing the RTI \eqref{KAineq} in terms of the geometrical parameters gives
\bea
{\cal I}
=
\frac{
	a^2\left[(a^2+r_+^2)(r_+^2+\ell^2)+q^2\ell^2\right]
}{
	a^2q^2\ell^2+
	(a^2+r_+^2)
	\left[a^2(\ell^2-r_+^2)+2r_+^2\ell^2\right]
}
<1.
\eea
For the fraction to be smaller than unity, its numerator must be smaller than its denominator. Their difference simplifies to
\be
2r_+^2(a^2+r_+^2)(a^2-\ell^2)<0,
\ee
which yields $ |a|<\ell $.
This is precisely the standard regularity condition required for a nonsingular KN-AdS geometry and coincides with the corresponding condition for Kerr-AdS.

Since, in the present convention, the angular momentum and mass satisfy $J=aM$, the condition $|a|<\ell$ can equivalently be expressed in terms of the conserved quantities as $ |J|<M\ell $. Thus, for the KN-AdS black hole, the proposed thermodynamic inequality reduces to the same bound in the $(M,J,\ell)$ parameter space as for Kerr-AdS, with the charge parameter $q$ dropping out of the final constraint.

The existence of an event horizon imposes an additional condition on the mass parameter,
\be
m\geq m_{\rm ext}(a,q,\ell),
\ee
where
\be
m_{\rm ext}
=
r_{\rm ext}
\left(
1+\frac{a^2+2r_{\rm ext}^2}{\ell^2}
\right),
\qquad
r_{\rm ext}^2
=
\frac{
	\sqrt{(a^2+\ell^2)^2+12\ell^2(a^2+q^2)}
	-(a^2+\ell^2)
}{6}.
\ee
Thus, the regularity condition $|a|<\ell$ and the horizon-existence condition $m\geq m_{\rm ext}$ play distinct roles: the former guarantees the absence of the rotational-coordinate singularity associated with $\Xi=1-a^2/\ell^2$, whereas the latter guarantees the existence of an event horizon. In the uncharged limit $q\to0$, the above extremality relations reduce to those of the Kerr-AdS black hole.

For the charged KS-AdS black hole, the relevant identity is given by Eq. \eqref{KS-AdSiden}, which has the same structure as the KN-AdS identity, with $M$ and $V$ replaced by the thermodynamic quantities $\widetilde M$ and $\widetilde V$, respectively. Therefore, the preceding argument establishing the existence of a positive root, and hence the validity of the proposed inequality, applies directly to the charged KS-AdS case as well.

It is instructive to examine the same inequality in terms of the geometrical parameters. Using the thermodynamic quantities given in the Appendix \ref{KSAblackhole}, together with the horizon equation $\Delta_r(r_+)=0$, we obtain
\be
{\cal I}
=
\frac{
	a^2\ell^2\left[\ell^2+r_+(r_++2b)\right]
}{
	(a^2+\ell^2)
	\left[
	a^2(\ell^2-r_+^2)
	+2\ell^2r_+(b+r_+)
	\right]
}<1,
\ee
for the inequality \eqref{KSAineq}, which implies
\be
a^4(\ell^2-r_+^2)
+2\ell^4r_+(r_++b)>0.
\ee
Using $b=q^2/(2m)$, this condition can equivalently be written as
\be\label{KSAineq}
a^4(\ell^2-r_+^2)
+2\ell^4r_+^2
+\frac{q^2\ell^4r_+}{m}>0.
\ee
For the standard KS-AdS branch, $|a|<\ell$, with $m>0$, $q^2\geq0$, and $r_+>0$, this inequality is automatically satisfied. Indeed, for $r_+\leq\ell$, all terms in \eqref{KSAineq} are non-negative and the second term is strictly positive. For $r_+>\ell$, the condition $|a|<\ell$ implies $a^4<\ell^4$, and consequently
\be
a^4(\ell^2-r_+^2)
>
-\ell^4(r_+^2-\ell^2),
\ee
so that
\be
a^4(\ell^2-r_+^2)
+2\ell^4r_+^2
+\frac{q^2\ell^4r_+}{m}
>
\ell^4(\ell^2+r_+^2)
+\frac{q^2\ell^4r_+}{m}>0.
\ee
Thus, on the standard regular KS-AdS branch, the regularity condition $|a|<\ell$ guarantees that the RTI is strictly satisfied.

The condition for the existence of an event horizon is independently determined by
\be
m\geq m_{\rm ext}(a,q,\ell),
\ee
where
\be
m_{\rm ext}
=
\frac{
	(r_{\rm ext}^2+2br_{\rm ext}+a^2)
	\left(\ell^2+r_{\rm ext}^2+2br_{\rm ext}\right)
}{
	2\ell^2r_{\rm ext}
},
\ee
and $r_{\rm ext}$ is the positive real root of
\be
3r_{\rm ext}^{4}
+8br_{\rm ext}^{3}
+\left(a^2+4b^2+\ell^2\right)r_{\rm ext}^{2}
-a^2\ell^2=0.
\ee

Thus, for the charged KS-AdS solution, the proposed inequality is automatically satisfied on the standard regular branch. Unlike the KN-AdS case, however, its direct formulation in terms of the conserved quantities yields the nontrivial bound $ |J|<\frac{1}{2}\widetilde M\ell $, where $\widetilde M$ is the thermodynamic mass entering the first law. To the best of our knowledge, this bound has not previously been reported for the KS-AdS solution, either in the charged case or in the uncharged limit. The horizon-existence condition $m\geq m_{\rm ext}$ is logically independent and ensures the presence of an event horizon.

For charged and rotating black strings, obtaining a direct exact algebraic identity analogous to that of the KN-AdS black hole is considerably more involved because of the higher-degree polynomial equations involved. Nevertheless, assuming the validity of RTI, we show below that it leads to a physically consistent and nontrivial bound on the charge parameter, which smoothly reduces to the previously obtained result in the uncharged limit.

In this case, the RTI \eqref{Urbsineq} takes the form
\bea
{\cal I}=
\frac{9 a^2 \left(a^2+\ell ^2\right)
	\left(q^2 \ell ^2+r_+^4\right)}
{\left(3 a^2+2 \ell ^2\right)
	\left[3 a^2 \left(q^2 \ell ^2+r_+^4\right)
	-q^2 \ell ^4+3 r_+^4 \ell ^2\right]}
<1 .
\eea
For the physical branch, this condition can be written as
\bea\label{chbsb}
q<q_{\rm ext}
\sqrt{\frac{a^2+\ell^2}{3a^2+\ell^2}},
\eea
where \(q_{\rm ext}\) denotes the critical value of the charge parameter \cite{Bakhtiarizadeh:2025uks},
\be\label{extchargepara}
q_{\rm ext}=\frac{\sqrt{3}r_+^2}{\ell},
\ee
at which the Hawking temperature vanishes, the horizon becomes degenerate, and the black string reaches extremality. For nonzero rotation, \(a\neq0\), one has
\be
\frac{a^2+\ell^2}{3a^2+\ell^2}<1.
\ee
Consequently, the bound in Eq.~\eqref{chbsb} is strictly stronger than the extremality bound,
\be
q<q_{\rm ext}.
\ee
Thus, for \(a\neq0\), satisfaction of the RTI automatically implies that the charge parameter lies below its extremal value. The RTI therefore imposes a stronger upper bound on the charge parameter than the condition required for the existence of a nondegenerate horizon.

For the charged rotating black string, we can further examine the relation between the RTI and the bound \(|{\cal J}|<{\cal M}\ell\). Introducing
\begin{equation}
	j\equiv\frac{{\cal J}}{{\cal M}\ell},
	\qquad
	X\equiv\frac{a^2}{\ell^2},
\end{equation}
the thermodynamic relations give
\begin{equation}
	j^2=\frac{9X(1+X)}{(2+3X)^2}.
\end{equation}
Solving for \(X\) yields
\begin{equation}
	X=
	\frac{4j^2+\sqrt{9-8j^2}-3}
	{6(1-j^2)}.
	\label{XofMJ}
\end{equation}
Requiring \(X\) to be real and non-negative on the physical branch leads to the bound \(|{\cal J}|<{\cal M}\ell\). Thus, this bound follows directly from the parameterization of the rotating black string.

The RTI imposes an additional condition. For the charged rotating black string, the RTI can be expressed in terms of the geometric parameters as given in Eq.~(\ref{chbsb}). Substituting the charge parameter \(q\) obtained from the horizon equation, \(f(r_+)=0\), and eliminating the metric parameters associated with the mass and rotation in favor of the thermodynamic mass and angular momentum, gives
\begin{equation}
	r_+^3>
	8\pi{\cal M}\ell^3
	\frac{1+3X}{(2+3X)^2}.
	\label{rplusRTI}
\end{equation}
This result shows that, unlike the uncharged case, the RTI in the charged case cannot be reduced to a condition involving only \({\cal M}\), \({\cal J}\), and \(\ell\). Even after eliminating the rotation parameter \(a\) and the mass parameter \(m\), the horizon radius \(r_+\), or equivalently the charge parameter \(q\), remains as an independent quantity. Therefore, the bound \(|{\cal J}|<{\cal M}\ell\) and the RTI encode distinct constraints and should not be regarded as equivalent forms of the same condition.

As a consistency check, in the non-rotating limit \(a\to0\), the condition \eqref{chbsb} reduces to \(q<q_{\rm ext}\), coinciding with the usual non-extremality bound. In the uncharged limit \(q\to0\), the condition \eqref{chbsb} reduces to
\be
3\left(1+\frac{a^2}{\ell^2}\right)>0,
\ee
which is automatically satisfied for real \(a\) and \(\ell\neq0\). Hence, in the \((m,a,\ell)\) parameterization, the RTI imposes no additional restriction on the rotation parameter. This is consistent with the uncharged rotating black string, for which the proposed inequality is identically satisfied throughout the allowed finite range of \(a\). In contrast, when expressed in terms of the conserved quantities, the same uncharged limit reproduces the previously obtained nontrivial bound \eqref{genonsing1}, $ |{\cal J}|<{\cal M}\ell $. Therefore, the charged result passes the expected consistency check: as \(q\to0\), the constraint becomes trivial in the metric-parameter space \((m,a,\ell)\), while it reduces to the nontrivial thermodynamic bound in the conserved-quantity space \(({\cal M},{\cal J},\ell)\).

The horizon-existence condition for the charged rotating black string is determined independently by the metric function. In particular, the mass parameter must satisfy
\begin{equation}
	m\geq m_{\rm ext}(q,\ell),
\end{equation}
where
\begin{equation}
	m_{\rm ext}
	=
	\frac{4r_{\rm ext}^3}{\ell^2},
	\qquad
	r_{\rm ext}
	=
	\left(\frac{q^2\ell^2}{3}\right)^{1/4}.
\end{equation}
Although the black string is rotating, the metric function \(f(r)\) is independent of the rotation parameter \(a\). Consequently, the condition for horizon existence, including the extremality bound, is independent of \(a\). This is in contrast to the constraint obtained from the proposed inequality, which depends explicitly on \(a\). Therefore, the RTI imposes a genuinely additional restriction on the charge parameter that is not simply equivalent to the usual horizon-existence or extremality condition.

For our next example, we consider the charged, rotating AdS C-metric. We begin with the thermodynamic inequality (\ref{C-metiden}). Assuming the physical condition $ M>0 $, we divide the inequality by $ M^2 $ and rewrite it as
\begin{equation}\label{haC-metric}
	A^3\leq36\pi V^2
	\left[
	1-\left(\frac{4\pi J^2}{3MV}\right)^2
	\right].
\end{equation}
Since a physical horizon must have a strictly positive area, $ A>0 $, compatibility of Eq.~(\ref{haC-metric}) with a positive horizon area requires the right-hand side to be strictly positive. This, in turn, yields the RTI \eqref{KAineq}. Hence, for the charged, rotating AdS C-metric, the RTI constitutes a necessary condition for the thermodynamic relation (\ref{haC-metric}) to admit a positive horizon area. It is worth emphasizing, however, that this argument establishes only a necessary thermodynamic condition and should not be interpreted as an independent proof of the existence of an event horizon. The actual existence of a horizon must still be established from the horizon equation of the C-metric. In particular, when $ {\cal I}=1 $, the right-hand side of Eq.~(\ref{haC-metric}) vanishes, implying $ A^3\leq0 $. Since $ A>0 $ for a physical horizon, no admissible positive-area solution exists in this case. Thus, saturation of the RTI is incompatible with a physical horizon of nonzero area.

It is worth emphasizing that the inequality does not merely reduce to the horizon-existence condition. Indeed, upon using the horizon equation to eliminate the mass parameter and substituting the thermodynamic quantities given in Appendix \ref{craC-metric}, the RTI \eqref{KAineq} takes the form
\begin{equation}
	1-{\cal I}
	=
	\frac{\mathcal{N}}{\mathcal{D}}>0,
\end{equation}
where
\begin{align}
	\mathcal{N}={}&
	2K^2(1+a^2A^2)
	r_+^2(r_+^2+a^2)\Xi
	\nonumber\\
	&+
	(1-A^2r_+^2)F
	\left[
	a^2(K^2-\Xi^2)
	+
	K^2A^2\ell^4\Xi^2
	\right],
\end{align}
and 
\begin{align}
	\mathcal{D}={}&
	2K^2(1+a^2A^2)
	r_+^2(r_+^2+a^2)\Xi
	\nonumber\\
	&+
	(1-A^2r_+^2)F
	K^2
	\left(
	a^2+A^2\ell^4\Xi^2
	\right).
\end{align}
with
\begin{equation}
	F=
	(1-A^2r_+^2)(r_+^2+a^2+q^2)
	+\frac{r_+^2(r_+^2+a^2)}{\ell^2}.
\end{equation}

It follows from the thermodynamic quantities that, on the physical branch, \(r_+>0\) is required for a physical horizon, while the condition \(0<Ar_+<1\) ensures that the black hole horizon lies within the physical region and does not intersect the conformal boundary. Moreover, the condition $ 1-A^2\ell^2\Xi>0 $ is required for \(\alpha\) to be real. Together with the positivity of the mass \(M\), this condition implies \(\Xi>0\). Consequently, $ F>0 $.
 Moreover, non-negative string tensions imply
$K\geq\Xi+2mA\geq\Xi$ for $A\geq0$ and $m>0$. Hence,
$\mathcal{N}>0$ and $\mathcal{D}>0$, yielding the RTI.

Thus, for the accelerating black holes considered here, the RTI is established upon imposing the physicality conditions of the solution, including the horizon and slow-acceleration conditions, together with the non-negativity of the string tensions. This demonstrates that the RTI contains information beyond the horizon equation alone, although its validity here relies on these physicality conditions.

The condition for the existence of a black hole horizon is
\begin{equation}
	m\geq m_{\rm ext}(a,q,\ell,A),
	\label{C-metric-horizon-condition}
\end{equation}
where
\begin{equation}
	m_{\rm ext}(a,q,\ell,A)
	=
	\frac{r_{\rm ext}^2+a^2+q^2}{2r_{\rm ext}}
	+
	\frac{r_{\rm ext}(r_{\rm ext}^2+a^2)}
	{2\ell^2(1-A^2r_{\rm ext}^2)}.
	\label{C-metric-mext}
\end{equation}
and $r_{\rm ext}$ denote the largest positive root of 
\begin{align}
	0={}&
	A^2(A^2\ell^2-1)r_{\rm ext}^{6}
	\nonumber\\
	&+
	\left[
	3+A^2a^2-2A^2\ell^2
	-A^4\ell^2(a^2+q^2)
	\right]r_{\rm ext}^{4}
	\nonumber\\
	&+
	\left[
	a^2+\ell^2
	+2A^2\ell^2(a^2+q^2)
	\right]r_{\rm ext}^{2}
	-\ell^2(a^2+q^2).
	\label{C-metric-rext}
\end{align}

We emphasize, however, that the existence of a positive value of $ A $ satisfying Eq.~(\ref{haC-metric}) should not by itself be identified with the existence of a regular event horizon of the C-metric. The latter requires, independently, an appropriate root structure of the horizon equation, together with the corresponding regularity conditions on the conical deficits and the restriction to the relevant acceleration regime. Therefore, the above argument establishes the necessity of the RTI for a positive horizon area compatible with the thermodynamic identity, but does not by itself establish sufficiency for the existence of a physical event horizon.

We now turn to rotating black holes with pairwise-equal charges in \(D=4\) gauged supergravity. These solutions, characterized by two \(U(1)\) charges, were first presented in \cite{Chong:2004na} and subsequently studied from a thermodynamic perspective in \cite{Cvetic:2010jb,Cvetic:2005zi}. By applying the relations given in Appendix \ref{pefcblackhole}, the RTI can be rewritten in terms of the geometric parameters of the solution as

\bea
{\cal I}&=&a^2 \left[\ell^2 \left[a^2+r_+ (2 (q_1+q_2)+r_+)\right]+r_1 r_2 \left(a^2+r_1 r_2\right)\right]\times\nn\\&&\left[a^2 \left(a^2+r_1 r_2\right) \left[r_1 r_2-2 r_+ (q_1+q_2+r_+)\right]\right.\nn\\&&\left.+\ell^2 \left\{a^4+a^2 r_+ \left[4 (q_1+q_2)+3 r_+\right]+2 r_1 r_2 r_+ (q_1+q_2+r_+)\right\}\right]^{-1}<1,
\eea
where $ r_1 $ and $ r_2 $ are defined in Eq.~\eqref{Pairwisepara}. The horizon equation has also been used to eliminate the mass parameter \(m\). The RTI can then be verified directly by requiring the numerator of the resulting fraction to be smaller than its denominator. Equivalently, the difference between the numerator and denominator must be negative, \ie
\be
r_+ (q_1+q_2+r_+)\left(a^2+r_1 r_2\right)\left(a^2-\ell^2\right)<0.
\ee
For the physical parameter range, the first three factors are positive. Consequently, the inequality reduces to (\ref{genonsing}) (since $ J=aM $, this may also be written as \eqref{genonsing1}). That is the regularity condition associated with Pairwise-equal four-charge AdS black hole in gauged supergravity.

The existence of a horizon requires the appropriate mass parameter to lie in the black-hole region,
\be
m\geq m_{\rm ext}(a,q_1,q_2,\ell),
\ee
where
\be
m_{\rm ext}
=
\frac{1}{2r_{\rm ext}}
\left[
r_{\rm ext}^{2}+a^{2}
+
\frac{
	(r_{\rm ext}+2q_1)(r_{\rm ext}+2q_2)
	\left[
	(r_{\rm ext}+2q_1)(r_{\rm ext}+2q_2)+a^{2}
	\right]
}{\ell^{2}}
\right],
\ee
and \(r_{\rm ext}\) is the positive real root of the extremality equation
\be
	3r_{\rm ext}^{4}
	+8(q_1+q_2)r_{\rm ext}^{3}
	+\left[
	\ell^2+a^2
	+4\left(q_1^2+4q_1q_2+q_2^2\right)
	\right]r_{\rm ext}^{2}
	-\left[
	a^2\ell^2
	+4a^2q_1q_2
	+16q_1^2q_2^2
	\right]
	=0.
\ee

Therefore, for this solution, the RTI is consistent with and reduces to the usual rotational regularity condition. However, it is not, by itself, a sufficient condition for the existence of an event horizon. This distinction is important because the horizon-existence condition contains additional dependence on the charge parameters \(q_1\) and \(q_2\), whereas the resulting thermodynamic bound, $ |J|<M\ell $, is independent of the charge parameters.

\section{Super-entropic black holes}\label{su}

We next consider the charged super-entropic black hole introduced in Ref.~\cite{Hennigar:2014cfa}. These solutions are obtained from the KN-AdS black hole by taking the ultraspinning limit \cite{Hennigar:2014cfa,Hennigar:2015cja}, in which the rotation parameter approaches the AdS length, \(a\to\ell\), while an appropriate rescaling and compactification of the azimuthal coordinate are performed. This procedure leads to black holes with noncompact horizon geometry but finite horizon area.

Using the compactification condition (\ref{psiC}), the area of the orthogonal two-dimensional surface is $ \omega_2=2\mu $. Consequently, the isoperimetric ratio becomes \cite{Hennigar:2014cfa}
\be\label{suIR}
\mathcal{R}
=
\left(\frac{r_+A}{2\mu}\right)^{1/3}
\left(\frac{2\mu}{A}\right)^{1/2}
=
\left(\frac{r_+^2}{r_+^2+\ell^2}\right)^{1/6}<1.
\ee
Thus, the charged super-entropic black holes necessarily violate the RII, providing a nontrivial test of the proposed thermodynamic inequality in a class of rotating AdS black holes with noncompact horizon topology.

Using the thermodynamic relations given in Appendix~\ref{SUKNblackhole} and eliminating the geometric parameters, we obtain the following exact algebraic identity:
\begin{equation}\label{SUidentity}
	A^2M^2
	\left(
	24\pi J^2MV-9M^2V^2-A^2J^2
	\right)
	=
	16\pi^2J^2Q^2
	\left(
	9M^2V^2+A^2J^2
	\right).
\end{equation}
In the uncharged limit, $ Q=0 $, this identity reduces to
\begin{equation}\label{SUidentityuncharged}
	24\pi J^2MV-9M^2V^2-A^2J^2=0.
\end{equation}
Solving for the horizon area gives
\begin{equation}
	A^2
	=
	24\pi MV
	\left(
	1-\frac{3MV}{8\pi J^2}
	\right)=24\pi MV\left(
	1-\frac{1}{2 {\cal I}}
	\right).
\end{equation}
Therefore, requiring a positive horizon area, \(A>0\), leads to
\begin{equation}\label{suKNAineq}
	{\cal I}>\frac{1}{2}.
\end{equation}
This result is qualitatively different from the rotational thermodynamic inequality  obtained for the KN-AdS black hole, for which the corresponding condition is \({\cal I}<1\). Thus, for the super-entropic KN-AdS black hole, positivity of the horizon area leads to a lower bound, \({\cal I}>1/2\), on the rotational thermodynamic ratio rather than an upper bound. This difference appears to originate from the special structure of the ultraspinning limit, in particular from the relation \(J=M\ell\), which is characteristic of the super-entropic geometry. Consequently, the rotational thermodynamic bound for super-entropic black holes is not simply a direct extension of the corresponding bound for the regular KN-AdS family.

Interestingly, the rotational thermodynamic inequality  obtained above is directly related to the isoperimetric ratio of the super-entropic black hole. Indeed, using the thermodynamic and geometric relations of Appendix~\ref{SUKNblackhole}, one finds
\begin{equation}
	\mathcal{R}^6=\frac{1}{2{\cal I}},
\end{equation}
or equivalently,
\begin{equation}
	\mathcal{R}^6{\cal I}=\frac{1}{2}.
\end{equation}
Consequently, the condition required by the positivity of the horizon area, $
\mathcal{I}>1/2 $, is equivalent to
\begin{equation}
	\mathcal{R}<1.
\end{equation}
Thus, for the super-entropic black hole, the RTI is precisely consistent with the violation of the reverse isoperimetric inequality. Rather than restoring the standard reverse isoperimetric bound \(\mathcal{R}\geq1\), the thermodynamic condition naturally leads to the opposite inequality for this class of solutions.

It is useful to rewrite the RTI of KN-AdS super-entropic black hole \eqref{suKNAineq} in terms of the geometric parameters of the solution and examine whether it leads to any additional constraint on these parameters.

In this limit, the horizon becomes noncompact while retaining a finite area, resulting in a geometry that is qualitatively distinct from the regular Kerr-AdS family with compact horizons. Using the mass parameter obtained from the horizon equation, together with the thermodynamic quantities listed in Appendix~\ref{SUKNblackhole}, the RTI for KN-AdS super-entropic black hole \eqref{suKNAineq} takes the form
\be\label{suinequality}
{\cal I}
=
\frac{(r_+^2+\ell^2)^2+q^2\ell^2}
{2r_+^2(r_+^2+\ell^2)}>\frac{1}{2} .
\ee
For the uncharged solution this reduces to
\be
\frac12\left(1+\frac{\ell^2}{r_+^2}\right)>\frac{1}{2}.
\ee
For $ r_+>0 $ and $ \ell>0 $, this inequality is manifestly satisfied.

For the charged solutions, the inequality \eqref{suinequality} instead yields
\be
\ell^2\left(r_+^2+\ell^2+q^2\right)>0,
\ee
which is automatically satisfied for real \(q\), \(r_+>0\), and \(\ell^2>0\). Thus, in the charged case, the condition \({\cal I}>1/2\) imposes no additional restriction on the charge parameter.

The location of the event horizon, $r_+$, is determined by the largest root of $\Delta_{r}(r_+)=0$. Horizons exist provided that the mass parameter satisfies \cite{Hennigar:2014cfa}
\be\label{suenext}
m \ge m_{\rm ext},
\ee
where
\be
m_{\rm ext}=2r_0\left(\frac{r_0^2}{\ell^2}+1\right)\, ,
\qquad
r_0^2 \equiv \frac{\ell^2}{3}
\left[
-1+\left(4+\frac{3q^2}{\ell^2}\right)^{1/2}
\right].
\ee
The equality in Eq. (\ref{suenext}) corresponds to the extremal black hole. 

\section{Reverse isoperimetric inequality}\label{rev}

In the Letter \cite{Amo:2023bbo}, the authors proposed novel thermodynamic inequalities for charged, stationary, and asymptotically AdS rotating black holes. Their approach is based on extending the reverse isoperimetric inequality (RII) \cite{Cvetic:2010jb} to rotating spacetimes. The RII conjectures that, for a black hole in \(D\) dimensions with horizon area \(A\) and thermodynamic volume \(V\), the isoperimetric ratio
\be\label{IR}
{\cal R}\equiv
\left(\frac{V}{V_0}\right)^{1/(D-1)}
\left(\frac{A_0}{A}\right)^{1/(D-2)},
\ee
satisfies $ {\cal R}\geq1 $.\footnote{Here, \(A_0\) and \(V_0\) denote the area and volume of a unit surface of constant \((t,r)\). For a sphere, \(A_0=\omega_{D-2}=2\pi^{(D-1)/2}/\Gamma[(D-1)/2]\) and \(V_0=\omega_{D-2}/(D-1)\).}
Physically, the RII states that, at fixed thermodynamic volume, the Schwarzschild-AdS black hole has the maximum possible entropy and hence the maximum horizon area. Equivalently, one may write
\be
A(V)\leq A_{\rm Schw}(V),
\ee
where \(A_{\rm Schw}(V)\) denotes the area of the Schwarzschild-AdS black hole with the same thermodynamic volume \cite{Amo:2023bbo}.

For rotating black holes, Ref.~\cite{Amo:2023bbo} proposed a strong refined reverse isoperimetric inequality (RRII). In four dimensions, for a stationary, asymptotically AdS black hole with mass \(M\), angular momentum \(J\), and thermodynamic volume \(V\), it takes the form
\be\label{RRII}
A(M,J,V)\leq A_{\rm Kerr}(M,J,V),
\ee
where \(A_{\rm Kerr}\) is the horizon area of the Kerr-AdS black hole with the same values of \(M\), \(J\), and \(V\). It is useful to introduce the function
\be
f(M,A,V,J)\equiv
36\pi M^2V^2-M^2A^3-64\pi^3J^4,
\ee
which vanishes identically for the Kerr-AdS solution, as discussed in Eq.~(\ref{K-AdSidentity}). In terms of this function, the refined inequality (\ref{RRII}) can be expressed as
\be
f(M,A,V,J)\geq0.
\ee

The refined inequality can also be expressed in terms of the isoperimetric ratio as \cite{Amo:2023bbo}
\be
{\cal R}\geq
\left[
1-\left(\frac{4\pi J^2}{3MV}\right)^2
\right]^{-1/6}.
\ee
In the nonrotating limit, \(J\to0\), the Kerr-AdS reference solution reduces to Schwarzschild-AdS, and the conventional RII, \({\cal R}\geq1\), is recovered. Moreover, the bound is saturated by the Kerr-AdS solution.

The role of the thermodynamic inequality proposed in this work becomes clear from the right-hand side of the above relation. If the RTI \eqref{KAineq} holds, then
\be
{\cal R}_{\rm Kerr-AdS}=
\left[
1-{\cal I}^2
\right]^{-1/6}\geq1.
\ee
Thus, the RTI ensures that the Kerr-AdS reference area entering the refined bound is real and positive and, consequently, that the right-hand side of the refined inequality is consistent with the conventional RII.

It is important to emphasize, however, that the RII does not follow from RTI alone. Rather, it follows from combining the RTI with the refined area bound (\ref{RRII}). The latter is taken as an input from Ref.~\cite{Amo:2023bbo} and is not independently established in the present work. Accordingly, the RII derived below is conditional on the validity of the refined entropy bound.

The two inequalities therefore play distinct roles. The RTI constrains the admissible thermodynamic parameter space and guarantees the consistency of the Kerr-AdS reference solution, while the refined inequality provides the additional relation required to compare the area of a general rotating AdS black hole with that of the corresponding Kerr-AdS solution. When these two ingredients are combined, the RTI is consistent with the RII for the class of regular compact-horizon solutions considered above. For super-entropic black holes, however, the situation is qualitatively different: the RII is violated, and we find that the RTI is violated consistently with this super-entropic behavior.

Accordingly, the conclusions concerning the RII obtained in this work are conditional on the refined entropy bound established in Ref.~\cite{Amo:2023bbo}. In particular, we do not claim that the RTI by itself implies the RII. Rather, our analysis identifies an additional thermodynamic constraint involving \(M\), \(J\), and \(V\), and, when combined with the refined entropy bound of Ref.~\cite{Amo:2023bbo}, investigates its consequences for the RII. Our purpose is therefore not to rederive the entropy bound of Ref.~\cite{Amo:2023bbo}, but to study the independent thermodynamic inequality proposed here and its implications for rotating AdS black holes with different horizon topologies.

\section{Higher-dimensional extension of the inequality}\label{ext}

In dimensions greater than four, a direct analogue of the four-dimensional derivation based on an algebraic identity for \(A(M,J,V)\) is not generally available, since eliminating the horizon and rotation parameters in favor of the thermodynamic variables typically leads to higher-degree polynomial relations. We therefore do not claim an independent derivation of the higher-dimensional RTI based solely on the positivity of the horizon area.

Instead, we consider the intermediate RRII introduced in Ref. \cite{Amo:2023bbo} obtained from the higher-dimensional Kerr-AdS thermodynamic relations. These inequalities were obtained by introducing the dimensionless quantity
\begin{equation}
	z=
	\frac{1+r_+^2/\ell^2}{r_+^2}
	\sum_i\frac{a_i^2}{\Xi_i}.
\end{equation}
The Kerr-AdS thermodynamic volume and angular momenta can be combined into a relation between $z$, $M$, $J_i$, and $V$. The isoperimetric ratio can then be expressed as a monotonic function of $z$, with separate expressions for even and odd spacetime dimensions. Then, using the corresponding Kerr-AdS relations between $M$, $J_i$, and $a_i$, one obtains the intermediate RRII in $ D $ dimensions. Assuming that the RII continues to hold for rotating black holes in arbitrary dimensions, the corresponding intermediate bounds lead to the following RTI depending on whether the spacetime dimension is even or odd and on whether the black hole carries a single nonzero angular momentum or multiple angular momenta:
\bea
\text{Even $ D $:}\; \begin{cases}\label{evendimension}
	\frac{2\pi(D-2)J_{\rm min}^2}{(D-1)MV}<1, & \text{for angular momenta $ J_i $ with $ J_{\rm min}={\rm min}\{ \left|J_i\right| \} $},\\
	\frac{8\pi J^2}{(D-1)(D-2)MV}<1, & \text{for a single non-zero angular momentum $ J $},
\end{cases}
\eea
and
\bea
\text{Odd $ D $:}\; \begin{cases}\label{odddimension}
	\frac{2\pi J_{\rm min}^2}{MV}<1, & \text{for angular momenta $ J_i $ with $ J_{\rm min}={\rm min}\{ \left|J_i\right| \} $},\\
	\frac{4\pi J^2}{(D-1)(D-2)MV}<1, & \text{for a single non-zero angular momentum $ J $}.
\end{cases}
\eea

For completeness, the analytical derivation of these relations is provided in Appendix~\ref{Ddimensionalineq}. The nontrivial content of the higher-dimensional proposal lies instead in the conjecture that the same thermodynamic combination provides a necessary condition for rotating, asymptotically AdS black holes beyond the Kerr-AdS family, provided that the intermediate RRII is assumed. In this sense, the higher-dimensional relations should be regarded as a conjectural extension rather than as a theorem established solely from the Myers-Perry-AdS solution.

As a further higher-dimensional test, we consider the singly spinning thin AdS black rings with horizon topology \(S^1\times S^{D-3}\), constructed perturbatively in \cite{Caldarelli:2008pz}. Unlike the Myers-Perry-AdS solutions considered above in formulating the higher-dimensional conjecture, these black rings provide a nontrivial test of the conjecture because of their nonspherical horizon topology. The corresponding thermodynamic quantities are given in Appendix \ref{ssablackring}. Introducing the dimensionless ring radius \(\mathsf{R}=R/\ell\), these quantities yield

\begin{equation}
	\frac{J^2}{MV}
	=
	\frac{D-1}{8\pi(D-2)}
	\frac{
		[1+(D-2)\mathsf{R}^2]
		[D-3+(D-2)\mathsf{R}^2]
	}{
		(1+\mathsf{R}^2)^2
	},
\end{equation}
for a thin AdS black ring with a single nonvanishing angular momentum \(J\). Consequently, for even \(D\), the RTI takes the form
\begin{equation}
	{\cal I}_{\rm even~D}=\frac{8\pi J^2}{(D-1)(D-2)MV}
	=
	\frac{
		[1+(D-2)\mathsf{R}^2]
		[D-3+(D-2)\mathsf{R}^2]
	}{
		(D-2)^2(1+\mathsf{R}^2)^2
	}<1,
\end{equation}
where
\begin{equation}
	1-
	\frac{
		[1+(D-2)\mathsf{R}^2]
		[D-3+(D-2)\mathsf{R}^2]
	}{
		(D-2)^2(1+\mathsf{R}^2)^2
	}
	=
	\frac{
		(D-2)^2\mathsf{R}^2+D^2-5D+7
	}{
		(D-2)^2(1+\mathsf{R}^2)^2
	}>0
\end{equation}
for $D\geq5$. For odd $D$, the proposed inequality similarly reduces to
\begin{equation}
	{\cal I}_{\rm odd~D}=\frac{4\pi J^2}{(D-1)(D-2)MV}
	=
	\frac{1}{2}
	\frac{
		[1+(D-2)\mathsf{R}^2]
		[D-3+(D-2)\mathsf{R}^2]
	}{
		(D-2)^2(1+\mathsf{R}^2)^2
	}<\frac12<1.
\end{equation}
Thus, the RTI is satisfied throughout the entire range of the thin-ring solutions, irrespective of whether $ D $ is even or odd.

It is also useful to compare this result with the well-known angular-momentum bound for regular asymptotically AdS solutions [47], which, for solutions with a single (positive) angular momentum, is given by Eq. (\ref{genonsing1}). This bound is satisfied by the thin black rings and is approached in the limit \(\mathsf{R}\to\infty\) [48]. Indeed, from the thermodynamic quantities above one obtains
\begin{equation}\label{wellknownbound}
	\left(\frac{J}{M\ell}\right)^2
	=
	\frac{
		\mathsf{R}^2[1+(D-2)\mathsf{R}^2]
		[D-3+(D-2)\mathsf{R}^2]
	}{
		(D-2)^2(1+\mathsf{R}^2)^3
	}<1,
\end{equation}
where
\begin{equation}
	1-\left(\frac{J}{M\ell}\right)^2
	=
	\frac{
		(D-2)^2+(D-2)^2\mathsf{R}^2
		+(D^2-5D+7)\mathsf{R}^4
	}{
		(D-2)^2(1+\mathsf{R}^2)^3
	}>0.
\end{equation}
Moreover,
\begin{equation}\label{limit}
	\lim_{\mathsf{R}\rightarrow\infty}\frac{|J|}{M\ell}=1.
\end{equation}
Hence, the thin AdS black rings provide a nontrivial higher-dimensional test of our proposed inequality: it is satisfied for both even and odd dimensions. We emphasize, however, that in this example the thermodynamic inequality is not equivalent to the bound $|J|\leq M\ell$; the latter is an independent bound obeyed by the regular black ring solutions and is saturated only in the infinitely long ring limit.

Hence, the thin AdS black rings provide a nontrivial higher-dimensional test of the proposed inequality, which is satisfied throughout the entire range of the AdS thin black ring solutions for both even and odd dimensions. It is important to note, however, that the thermodynamic inequality is not algebraically equivalent to the angular-momentum bound \(|J|\leq M\ell\). Indeed, from Eq.~\eqref{wellknownbound}, one finds that the angular-momentum bound involves the same polynomial factor as the thermodynamic inequality, but with an additional prefactor \(\mathsf R^2/(1+\mathsf R^2)\). Therefore, the two inequalities are not algebraically equivalent. In addition to the different prefactor, the thermodynamic inequality contains a dimension-dependent normalization that differs between even and odd \(D\). Although both bounds are satisfied throughout the thin-ring family, they have distinct functional forms and exhibit different limiting behavior. In particular, while the ratio in Eq.~(\ref{limit}) approaches unity as $ \mathsf{R}\to\infty $, the proposed inequality approaches \(1\) for even \(D\) and \(1/2\) for odd \(D\). Thus, the saturation of the angular-momentum bound in the infinitely long-ring limit should not be identified with saturation of the proposed thermodynamic inequality.

\section{Conclusion}\label{conc}

In this work, we have investigated a rotational thermodynamic inequality (RTI) for stationary, asymptotically AdS rotating black holes. The proposed inequality involves the angular momentum, mass, and thermodynamic volume and provides a simple constraint on the thermodynamic parameter space of rotating AdS solutions. Our analysis shows that the same thermodynamic combination emerges naturally in a variety of geometries, although its precise physical interpretation is not universal and depends on the properties of the underlying solution.

For four-dimensional black holes, we first established the inequality from exact thermodynamic identities for several rotating AdS families. We then reformulated it in terms of the primary metric parameters and compared it with the conditions governing rotational regularity and horizon existence. For Kerr-AdS and several related rotating compact-horizon geometries, the resulting constraint reduces to the familiar rotational regularity condition. In particular, for Kerr-AdS and KN-AdS it is equivalent to the standard bound on the rotation parameter, while for KS-AdS it yields a corresponding bound expressed naturally in terms of the thermodynamic mass entering the first law. These results also demonstrate that the regularity condition associated with the rotation parameter and the condition for the existence of an event horizon are logically distinct. The latter is determined independently by the horizon equation and the corresponding extremality condition.

The analysis of rotating black strings provides a different perspective. For the uncharged rotating black string, the proposed inequality is automatically satisfied throughout the physical parameter space when expressed in terms of the metric parameters, although it can be written as a nontrivial bound in terms of the conserved quantities. For the charged rotating black string, on the other hand, the inequality gives an additional constraint on the charge parameter that is stronger than the usual extremality bound for nonzero rotation. This illustrates that the thermodynamic inequality need not simply reproduce an existing regularity condition and can instead impose an additional restriction on the physical parameters of a given geometry.

We have also examined the charged accelerating AdS C-metric. In this case, the thermodynamic relation implies that the proposed inequality is necessary for compatibility with a positive horizon area. A direct analysis in terms of the metric parameters further shows that the inequality follows from the physicality conditions of the accelerating solution and the positivity of the string tensions. We have emphasized that this argument does not, by itself, establish the existence of an event horizon: the latter requires an independent analysis of the horizon equation and the associated regularity conditions. This distinction is essential when interpreting the thermodynamic inequality in geometries with acceleration.

A qualitatively different behavior is found for super-entropic black holes. The ultraspinning limit changes the structure of the thermodynamic relation such that positivity of the horizon area leads to a lower, rather than an upper, bound on the rotational thermodynamic ratio. Moreover, this condition is directly related to the isoperimetric ratio and is consistent with the characteristic violation of the RII by super-entropic solutions. Thus, the super-entropic case demonstrates that the RTI is sensitive to the geometric structure and horizon topology and should not be regarded as a universal upper bound applicable without qualification to all rotating AdS black holes.

We have further clarified the relation between the RTI and the RII. The rotational thermodynamic inequality ensures that the Kerr-AdS reference quantity entering the refined reverse isoperimetric bound is well defined. However, the RII does not follow from the RTI alone. Its derivation requires the refined area bound established in Ref.~\cite{Amo:2023bbo} as an additional input. Accordingly, our results should be understood as establishing the thermodynamic constraint and examining its consequences when combined with the previously proposed refined entropy bound, rather than as an independent derivation of that bound.

Finally, we considered the higher-dimensional extension of the RTI. Since the direct four-dimensional algebraic construction does not generally extend to higher dimensions, the corresponding relations are formulated as a conjectural extension based on the higher-dimensional Kerr-AdS thermodynamic relations and the intermediate refined reverse isoperimetric bounds. As a nontrivial test beyond spherical horizon topology, we examined singly spinning thin AdS black rings. The RTI is satisfied throughout the thin-ring regime for both even and odd spacetime dimensions. We also showed that, although the RTI and the familiar angular-momentum bound are both satisfied by these solutions, they are not algebraically equivalent and exhibit different limiting behavior. In particular, saturation of the angular-momentum bound in the infinitely long-ring limit does not correspond to saturation of the proposed thermodynamic inequality.

Taken together, these results support the usefulness of the proposed thermodynamic combination as a constraint on rotating AdS black-hole solutions while also demonstrating that its physical interpretation depends on the geometry, matter content, and horizon topology. The examples considered here show that the RTI can reproduce known regularity conditions, provide genuinely additional restrictions, or change its character in ultraspinning geometries. A natural direction for future work is to investigate the proposed constraint for broader classes of higher-dimensional rotating solutions, including charged and multiply spinning black holes, as well as other nonspherical horizon topologies. Such analyses may help determine the extent to which the observed thermodynamic structure persists beyond the families considered in this work.

\providecommand{\href}[2]{#2}\begingroup\raggedright
\endgroup


\appendix

\section{Kerr-Newman-AdS black hole}\label{KNblackhole}

The metric of the KN-AdS black hole, expressed in Boyer-Lindquist-like coordinates, takes the form \cite{Carter:1968ks}
\bea\label{KN-AdSmetric} 
&\!\!\!\!\!\!\!\!\!\!&ds^2 = -\frac{\Delta_r}{\rho^2} \lp dt - \frac{a \sin^2\theta}{\Xi}d\phi \rp^2 + \frac{\rho^2}{\Delta_r} dr^2 + \frac{\rho^2}{\Delta_\theta} d\theta^2 + \frac{\sin^2\theta \Delta_\theta}{\rho^2}\lp adt - \frac{r^2+a^2}{\Xi} d\phi \rp^2,\nn\\
&\!\!\!\!\!\!\!\!\!\!&A = - \frac{q r}{\rho^2} \lp dt - \frac{a \sin^2\theta}{\Xi} d\phi \rp,
\eea
where
\bea\label{KNpara}
&&\rho^2 = r^2 + a^2 \cos^2\theta, \quad \Delta_\theta = 1 - \frac{a^2}{\ell^2} \cos^2\theta, \quad \Delta_r = (r^2+a^2)\left(1+\frac{r^2}{\ell^2}\right) - 2 m r + q^2, \nn\\ &&\Xi = 1 - \frac{a^2}{\ell^2},    
\eea
and the parameters $m$, $q$, and $a$ are related to the mass $M$, electric charge $Q$, and angular momentum $J$ of the KN-AdS black hole according to
\bea\label{KNMCA}
M = \frac{m}{\Xi^2},
\quad
Q = \frac{q}{\Xi},
\quad
J =aM= \frac{am}{\Xi^2}.
\eea
For the standard KN-AdS solution with a regular conformal boundary and the standard $2\pi$ periodicity of the azimuthal coordinate, one imposes $\Xi>0$, which is equivalent to $ |a|<\ell $.
The event horizon is located at $r=r_+$, where $r_+$ is the largest positive root of $\Delta_r=0$. Using $\Delta_r(r_+)=0$, the mass can be expressed as
\bea\label{KNmass}
M=\frac{(r_+^2+a^2)(r_+^2+\ell^2)+q^2\ell^2}{2r_+\ell^2\Xi^2}.
\eea
Moreover, the Bekenstein-Hawking entropy is
\be
S = \frac{\pi(r_+^2+a^2)}{\Xi}.
\ee

The extended thermodynamics of KN-AdS black holes was studied in \cite{Caldarelli:1999xj,Gunasekaran:2012dq}. Solving for $r_+$ in terms of $S$, substituting the resulting expression into Eq.~(\ref{KNmass}), and using Eq.~(\ref{KNMCA}) together with
\be\label{pressure}
P=-\frac{\Lambda}{8\pi}=\frac{3}{8\pi\ell^2},
\ee
the mass can be reexpressed in terms of the thermodynamic variables $(S,P,J,Q)$ as
\be\label{KNM}
M=\frac{1}{2}\sqrt{\frac{S}{\pi}\ls\lp1+\frac{\pi Q^2}{S}+\frac{8PS}{3}\rp^2+ \frac{4\pi^2J^2}{S^2}\lp1+\frac{8PS}{3}\rp\rs}.
\ee
The first law of thermodynamics in the extended phase space takes the form
\be\label{first}
dM=TdS+VdP+\Omega dJ+\Phi dQ,
\ee
where $T$, $V$, $\Omega$, and $\Phi$ denote the Hawking temperature, thermodynamic volume, angular velocity, and electric potential, respectively. Thus, the mass $M$ is naturally interpreted as the enthalpy of the black hole rather than its internal energy. The thermodynamic volume is consequently obtained from
\bea\label{KNthermovol}
V=\lp\frac{\pa M}{\pa P}\rp_{S,J,Q}
=\frac{2S^2}{3\pi M}
\lp1+\frac{\pi Q^2}{S}+\frac{8PS}{3}+\frac{2\pi^2J^2}{S^2}\rp.
\eea

Solving Eq.~(\ref{KNthermovol}) for $P$ and substituting the result into Eq.~(\ref{KNM}), while using the Bekenstein-Hawking relation $S=A/4$, allows the pressure to be eliminated in favor of the thermodynamic volume and the horizon area. One then obtains the following identity \cite{Amo:2023bbo}:
\be\label{KN-AdSiden}
36 \pi M^2 V^2 -M^2 A^3 -64 \pi^3 J^4
=16 \pi^2 Q^2 J^2 A.
\ee

\section{Kerr-Sen-AdS black hole}\label{KSAblackhole}

In the usual Boyer--Lindquist coordinates, the KS-AdS metric takes the form \cite{Wu:2020cgf}
\bea\label{KSAmet} 
&\!\!\!\!\!\!\!\!\!\!&ds^2 = -\frac{\Delta_r}{\rho^2} \lp dt - \frac{a \sin^2\theta}{\Xi}d\phi \rp^2 + \frac{\rho^2}{\Delta_r} dr^2 + \frac{\rho^2}{\Delta_\theta} d\theta^2 + \frac{\sin^2\theta \Delta_\theta}{\rho^2}\lp adt - \frac{r^2+2br+a^2}{\Xi} d\phi \rp^2,\nn\\
&\!\!\!\!\!\!\!\!\!\!&A = - \frac{q r}{\rho^2} \lp dt - \frac{a \sin^2\theta}{\Xi} d\phi \rp,
\eea
where
\bea\label{KSApara}
&&\rho^2 = r^2 +2br+ a^2 \cos^2\theta, \quad \Delta_r = (r^2+2br+a^2)\left(1+\frac{r^2+2br}{\ell^2}\right) - 2 m r, \nn\\ &&\Delta_\theta = 1 - \frac{a^2}{\ell^2} \cos^2\theta, \quad \Xi = 1 - \frac{a^2}{\ell^2}.   
\eea
Here, $b$ denotes the dilatonic parameter and is defined by
\begin{equation}
	b \equiv \frac{q^2}{2m},
\end{equation}
where $q$ and $m$ are related to the electric charge and mass parameters, respectively. In the limit $\ell\rightarrow\infty$, the metric reduces to the asymptotically flat Kerr-Sen black hole. In the non-rotating limit, $a=0$, it reduces to the Gibbons--Maeda--Garfinkle--Horowitz--Strominger solution. The Gibbons--Maeda solution was originally obtained in \cite{Gibbons:1987ps}, while the charged generalization was subsequently developed by Garfinkle, Horowitz, and Strominger \cite{Garfinkle:1990qj}. Four-dimensional KS-(A)dS black holes have also been studied from various perspectives, including their shadow properties \cite{Zhang:2021wda} and their thermodynamic behavior in the extended phase space \cite{Sharif:2021yis}.

The thermodynamic quantities associated with the KS-AdS solution (\ref{KSAmet}) in the extended phase space are given by \cite{Wu:2020cgf}
\bea\label{KSMCA}
&&M = \frac{m}{\Xi^2}, 
\quad 
Q = \frac{q}{\Xi}, 
\quad 
J =\frac{am}{\Xi^2},
\quad
T=\frac{(r_++b)\lp2r_+^2+4br_++\ell^2+a^2\rp-m\ell^2}{2\pi\lp r_+^2+2br_++a^2\rp\ell^2}, 
\nn\\
&&S=\frac{\pi\lp r_+^2+2br_++a^2\rp}{\Xi}, 
\quad
\Omega=\frac{a\Xi}{\lp r_+^2+2br_++a^2\rp}, 
\quad
\Phi=\frac{qr_+}{\lp r_+^2+2br_++a^2\rp},
\eea
where $r_+$ denotes the radius of the event horizon and the mass parameter $m$ is determined by the horizon condition $ \Delta_r(r = r_+) = 0 $.

These thermodynamic quantities satisfy the Bekenstein-Smarr relation
\be\label{Smarr}
M=2TS+2\Omega J+\Phi Q-2PV,
\ee
where the thermodynamic volume $V$, conjugate to the pressure $P$ defined in Eq.~(\ref{pressure}), is given by \cite{Wu:2020cgf}
\be
V=\frac{4\pi}{3\Xi}\lp r_++b\rp\lp r_+^2+2br_++a^2\rp
\ee
However, the corresponding first law takes the form
\be
dM=TdS+\Omega dJ+\Phi dQ-VdP+Jd\Xi/(2a).
\ee
which is a differential identity rather than the standard first law. This is because the thermodynamic quantities above are defined in a frame that rotates at infinity rather than in the non-rotating frame.

The KS-AdS solution can be transformed to the non-rotating frame at infinity by the coordinate transformation
\be
\phi=\phi-\frac{a}{\ell^2}t.
\ee
After transforming to this frame, the mass, angular velocity, and thermodynamic volume are modified according to \cite{Wu:2020cgf}
\be\label{restmassangthermovol}
\widetilde{M}=M+\frac{a}{\ell^2}J,\qquad{\widetilde \Omega}=\Omega+\frac{a}{\ell^2},\qquad{\widetilde V}=V+\frac{4\pi}{3}aJ.
\ee
In the non-rotating frame at infinity, the thermodynamic quantities satisfy both the standard first law and the corresponding Bekenstein-Smarr relation,
\bea\label{stfs}
&&d\widetilde{M}=TdS+{\widetilde \Omega} dJ+\Phi dQ-{\widetilde V}dP,\nn\\
&&\widetilde{M}=2TS+2{\widetilde \Omega} J+\Phi Q-2P{\widetilde V}.
\eea
Furthermore, both the differential first law and the integral Smarr relation follow from the Christodoulou-Ruffini-type squared-mass formula \cite{Wu:2020cgf}
\be\label{KSAM}
{\widetilde M}^2=\lp1+\frac{8PS}{3}\rp\ls\lp1+\frac{8PS}{3}\rp\frac{S}{4\pi}+\frac{\pi J^2}{S}+\frac{Q^2}{2}\rs. 
\ee

\section{ Charged and rotating AdS black string}\label{crablackstring}

The metric of an asymptotically AdS charged and rotating black string with cylindrical or toroidal horizon topology is given by \cite{Lemos:1994xp,Lemos:1995cm}
	\bea\label{met}
	&&ds^2=-f(r)\lp \Xi dt -a d\phi \rp ^2+\frac{1}{f(r)}dr^2+\frac{r^2}{\ell^4} \lp a dt -\Xi \ell^2 d\phi \rp ^2+\frac{r^2}{\ell^2}dz^2,\nn\\
	&&A=-\frac{q\Xi}{r}\lp dt-\frac{a}{\Xi}d\phi\rp,
	\eea
	where
	\be\label{sileq}
	f(r)=\frac{r^2}{\ell^2}-\frac{m}{r}+\frac{q^2}{r^2},\quad \Xi=\sqrt{1+\frac{a^2}{\ell^2}}.
	\ee
	Here, the integration constants $ m $ and $ q $ are proportional to the mass and charge of the black string, respectively. The constants $ a $ and $ \ell $, both having dimensions of length, represent the rotation parameter and the AdS radius, respectively. The relevant thermodynamic potentials per unit length of a stationary black string horizon with cylindrical topology are given by \cite{Dehghani:2002rr}
	\bea\label{bsMCA}
	{\cal M}=\frac{1}{16 \pi \ell}\lp 3 \Xi^2-1\rp m,\quad {\cal Q}=\frac{q \Xi}{4 \pi\ell},\quad
	{\cal J}=\frac{3}{16 \pi \ell} \Xi a  m,
	\eea
	where $ {\cal M} $, $ {\cal J} $ and $ {\cal Q} $ denote the mass, angular momentum, and electric charge per unit length of the black string, respectively. The event horizon is located at the largest root of the metric function determined by \ie $ f(r_+)=0 $. This yields the black string mass
	\be\label{bsm}
	{\cal M}=\frac{\lp 3 \Xi^2-1\rp\lp r_+^4+q^2\ell^2\rp}{16 \pi r_+ \ell^3}.
	\ee
	Furthermore, the entropy of black string per unit length of horizon is given by
	\be
	{\cal S}=\frac{ r_{+}^{2}\Xi}{4\ell}.
	\ee
	The remaining thermodynamic quantities per unit length are given by \cite{Dehghani:2002rr}
	\bea\label{thermopot}
	T=\frac{3 r_{+}^{4}-q^2 \ell^2}{4 \pi r_{+}^{3} \ell^2 \Xi},\qquad
	\Omega=\frac{a}{\Xi \ell^2},
	\eea
	
	The thermodynamics of asymptotically AdS charged and rotating black strings was extended to the extended phase space in \cite{Bakhtiarizadeh:2025uks}.
	
	Accordingly, the first law of black hole thermodynamics is modified \cite{Bakhtiarizadeh:2025uks}:
	\be\label{exfirstlaw}
	d{\cal M} = Td{\cal S} + \Omega d{\cal J} + \Phi d{\cal Q}+{\cal V}dP.
	\ee
	The extended first law is consistent with the corresponding integrated mass relation, namely, the Smarr relation
	\be\label{exsmarr}
	0 = T {\cal S} + \Omega {\cal J} -2 P {\cal V}.
	\ee
	In deriving the above relation, we have used Euler's theorem for homogeneous functions.
	
	The mass ${\cal M}$ expressed in terms of the thermodynamic variables $({\cal S},P,{\cal J},{\cal Q})$ is given by \cite{Bakhtiarizadeh:2025uks}
	\bea\label{mass}
	&&\!\!\!\!\!\!\!\!\!\!{\cal M} =\frac{1}{2^{3/4} \sqrt{3} \sqrt[4]{\pi } \sqrt[4]{P} \sqrt{\mathcal{S}} \left(32 \pi ^{3/2} \sqrt{2} \mathcal{J}^2 P^{3/2} \mathcal{S}+\varUpsilon \right)^{3/2}} \left[32 \pi ^{3/2} \sqrt{2} \mathcal{J}^2 P^{3/2} \mathcal{S} \varUpsilon +36864 P^4 \mathcal{S}^8\right.\nn\\&&\!\!\!\!\!\!\!\!\!\!\left.+55296 \pi  P^3 \mathcal{Q}^2 \mathcal{S}^6+32 \pi ^3 P \mathcal{S}^2 \left(64 \mathcal{J}^4 P^2+243 \mathcal{Q}^6\right)+31104 \pi ^2 P^2 \mathcal{Q}^4 \mathcal{S}^4+729 \pi ^4 \mathcal{Q}^8 \right],
	\eea
	where
	\bea\label{varUpsilon}
	&&\!\!\!\!\!\!\!\!\!\!\varUpsilon\equiv\left[110592 P^4 \mathcal{S}^8+165888 \pi  P^3 \mathcal{Q}^2 \mathcal{S}^6+32 \pi ^3 P \mathcal{S}^2 \left(64 \mathcal{J}^4 P^2+729 \mathcal{Q}^6\right)\right.\nn\\ &&\left.+93312 \pi ^2 P^2 \mathcal{Q}^4 \mathcal{S}^4+2187 \pi ^4 \mathcal{Q}^8\right]^{1/2}.
	\eea
	Using the first law, the thermodynamic volume, temperature, angular velocity, and electric potential can be expressed in terms of the thermodynamic variables $({\cal S},P,{\cal J},{\cal Q})$. In particular, the thermodynamic volume is given by \cite{Bakhtiarizadeh:2025uks}
	\bea\label{chthermovol}
	&&\!\!\!\!\!\!\!\!\!\!\mathcal {V}= \frac{\sqrt{32 \pi ^{3/2} \sqrt{2} \mathcal{J}^2 P^{3/2} \mathcal{S}+\varUpsilon }}{108\ 2^{3/4} \sqrt{3} \sqrt[4]{\pi } P^{5/4} \sqrt{\mathcal{S}} \left(8 P \mathcal{S}^2+3 \pi  \mathcal{Q}^2\right)^5}\times\nn\\&&\left[256 \pi ^{5/2} \sqrt{2} \mathcal{J}^2 P^{3/2} \mathcal{Q}^2 \mathcal{S} \varUpsilon +884736 P^5 \mathcal{S}^{10}+1216512 \pi  P^4 \mathcal{Q}^2 \mathcal{S}^8+580608 \pi ^2 P^3 \mathcal{Q}^4 \mathcal{S}^6\right.\nn\\&&\left.-8 \pi ^4 P \mathcal{Q}^2 \mathcal{S}^2 \left(2048 \mathcal{J}^4 P^2+729 \mathcal{Q}^6\right)+93312 \pi ^3 P^2 \mathcal{Q}^6 \mathcal{S}^4-2187 \pi ^5 \mathcal{Q}^{10}\right],
	\eea
	while the temperature is
	\bea\label{temp}
	&&\!\!\!\!\!\!\!\!\!\! T = -\frac{9 \sqrt{3} \left(\pi  \mathcal{Q}^2-8 P \mathcal{S}^2\right) \left(8 P \mathcal{S}^2+3 \pi  \mathcal{Q}^2\right)^3 \left(32 \pi ^{3/2} \sqrt{2} \mathcal{J}^2 P^{3/2} \mathcal{S} \varUpsilon +\varUpsilon ^2\right)}{2\ 2^{3/4} \sqrt[4]{\pi } \sqrt[4]{P} \mathcal{S}^{3/2} \varUpsilon  \left(32 \pi ^{3/2} \sqrt{2} \mathcal{J}^2 P^{3/2} \mathcal{S}+\varUpsilon \right)^{5/2}},
	\eea
	\bea
	\Omega= 
	\frac{16\ 2^{3/4} \pi ^{5/4} \mathcal{J} P^{5/4} \sqrt{\mathcal{S}}}{\sqrt{3} \sqrt{32 \pi ^{3/2} \sqrt{2} \mathcal{J}^2 P^{3/2} \mathcal{S}+\varUpsilon }},
	\eea
	and
	\bea
	&&\!\!\!\!\!\!\!\!\!\!\Phi = \frac {\sqrt[4]{2} \pi ^{3/4} \mathcal{Q} \sqrt{32 \pi ^{3/2} \sqrt{2} \mathcal{J}^2 P^{3/2} \mathcal{S}+\varUpsilon }}{27 \sqrt{3} \sqrt[4]{P} \sqrt{\mathcal{S}} \left(8 P \mathcal{S}^2+3 \pi  \mathcal{Q}^2\right)^5}\times  \nn\\&&\!\!\!\!\!\!\!\!\!\!\left[-64 \sqrt{2} \pi ^{3/2} \mathcal{J}^2 P^{3/2} \mathcal{S} \varUpsilon +110592 P^4 \mathcal{S}^8+165888 \pi  P^3 \mathcal{Q}^2 \mathcal{S}^6\right.\nn\\&&\left.+32 \pi ^3 P \mathcal{S}^2 \left(128 \mathcal{J}^4 P^2+729 \mathcal{Q}^6\right)+93312 \pi ^2 P^2 \mathcal{Q}^4 \mathcal{S}^4+2187 \pi ^4 \mathcal{Q}^8\right].
	\eea 
	Transforming back to the geometric variables $(r_+,a~({\rm or}~\Xi),q,\ell)$, one obtains the corresponding thermodynamic quantities in terms of the geometric parameters.
	
\section{Charged and rotating AdS C-metric}\label{craC-metric}	
	
The metric of the charged and rotating AdS C-metric is given by \cite{Anabalon:2018qfv}
	\bea
		ds^2 &\!\!\!\!\!=\!\!\!\!\!& \frac{1}{H^2} \left\{-\frac{f}{\Sigma} \left[\frac{dt}{\alpha} - a \sin^2 \theta \frac{d\phi}{K} \right]^2 + \frac{\Sigma}{f} dr^2 + \frac{\Sigma r^2}{h} d\theta^2 + \frac{h \sin^2\theta}{\Sigma r^2} \left[\frac{a dt}{\alpha} - (r^2+a^2) \frac{d\phi}{K}\right]^2 \right\},\nn\\ 
		\mathcal{A} &\!\!\!\!\!=\!\!\!\!\!& - \frac{q}{\Sigma r} \left[ \frac{dt}{\alpha} - a \sin^2\theta \frac{d\phi}{K} \right],
    \eea
where
    \bea
		&\!\!\!\!\!\!\!\!\!\!&f = \left(1-A^2 r^2\right) \left[1- \frac{2m}{r} + \frac{a^2+q^2}{r^2} \right]+ \frac{r^2+a^2}{\ell^2}, \quad \Sigma = 1 + \frac{a^2}{r^2} \cos^2 \theta, 
		\quad 
		H = 1 + A r \cos \theta,
		\nn\\&\!\!\!\!\!\!\!\!\!\!&h = 1 + 2 m A \cos \theta + \left[A^2(a^2+q^2) - \frac{a^2}{\ell^2} \right] \cos^2\theta \, .
	\eea
The parameter $A$ characterizes the acceleration of the spacetime. The event horizon is located at the largest root of $f(r_+)=0$. In the slow-acceleration regime, $A\ell<1$, which is the regime considered here, no acceleration horizon is present. 
	
The thermodynamic charges and potentials of the accelerating black hole are given by \cite{Anabalon:2018qfv}
\bea\label{AdSthermo}
&&M= \frac{m(\Xi+\frac{a^2}{\ell^2})(1-A^2 \ell^2\Xi)}{K\Xi\alpha(1+a^2A^2)}\,,\qquad
T= \frac{f'_+ r_+^2}{4\pi\alpha(r_+^2+a^2)}\,, \qquad
S=\frac{\pi(r_+^2+a^2)}{K(1-A^2r_+^2)}\,,\nn\\
&&Q= \frac{q}{K}\,,\qquad \Phi=\Phi_t=\frac{qr_+}{(r_+^2+a^2)\alpha}\,,\qquad
J =\frac{ma}{K^2}\,,\nn\\
&&\Omega=  \Omega_H-\Omega_\infty\,
=\left ( \frac{Ka}{\alpha(r_+^2+a^2)}\right ) -
\left ( -\frac{aK(1-A^2\ell^2\Xi)}{\ell^2\Xi \alpha(1+a^2A^2)}\right) \,,\nn\\
&&V = \frac{4\pi}{3K\alpha} \left [ \frac{r_+(r_+^2 + a^2)}{(1-A^2 r_+^2)^{2}}
+ \frac{m[a^2(1-A^2\ell^2 \Xi) + A^2 \ell^4 \Xi (\Xi+\frac{a^2}{\ell^2})]}{(1+a^2 A^2) \Xi} \right]\,,\nn\\
&&\lambda_\pm = \frac{r_+}{\alpha(1\pm Ar_+)} - \frac{m}{\alpha}
\frac{[\Xi + \frac{a^2}{\ell^2} +   \frac{a^2}{\ell^2} (1-A^2\ell^2 \Xi)]}{(1+a^2 A^2)\Xi^2}
\mp \frac{A \ell^2 (\Xi +  \frac{a^2}{\ell^2} )}{\alpha(1+a^2A^2)}\,,
\eea
where
\be
\Xi=1-\frac{a^2}{\ell^2}+A^2(q^2+a^2),
\ee
and the conical tensions $\mu_\pm$ are defined as 
\be\label{eq:deficits}
\mu_\pm =\frac{1}{4}\bigg[1-\frac{\Xi \pm 2mA}{K}\bigg]\,.
\ee
These quantities satisfy both the first law 
\be\label{flaw}
\delta M=T\delta S+\Phi \delta Q+\Omega \delta J-\lambda_+\delta \mu_+
-\lambda_-\delta \mu_- +V\delta P\,,
\ee
and the Smarr relation
\be\label{CSmarr}
M=2(TS+\Omega J-PV)+\Phi Q\,,
\ee
provided that
\be\label{alp-eq}
\alpha=\frac{\sqrt{(\Xi+\frac{a^2}{\ell^2})(1-A^2 \ell^2\Xi)}}{1+a^2A^2}\,.
\ee	
	
The extended thermodynamics of the charged and rotating AdS C-metric was analyzed in detail in \cite{Anabalon:2018qfv}. We also make use of results presented in \cite{Gregory:2019dtq}. For further details concerning the thermodynamic framework and its derivation, we refer the reader to these works.

We next review the derivation of the identity involving the thermodynamic variables of the charged and rotating AdS C-metric presented in Ref.~\cite{Amo:2023bbo}. While Eq.~(11) of \cite{Anabalon:2018qfv} provides the thermodynamic potentials in terms of the geometric parameters, we find it more convenient for our purposes to use Eqs.~(10) and (11) of \cite{Gregory:2019dtq}. The relevant expressions are
	\begin{align}
		M^2 &= \frac{\Delta_{\rm C} S}{4 \pi} \left[ \left(1 + \frac{\pi Q^2}{\Delta_{\rm C} S} + \frac{8 P S}{3 \Delta_{\rm C}} \right)^2  + \left(1 + \frac{8 P S}{3 \Delta_{\rm C}} \right)\left(\frac{4 \pi^2 J^2}{\Delta_{\rm C}^2 S^2} - \frac{3 C^2 \Delta_{\rm C}}{2 P S} \right)\right] \, ,
		\\ 
		V &= \frac{2 S^2}{3 \pi M} \left[\left(1 + \frac{\pi Q^2}{\Delta_{\rm C} S} + \frac{8 P S}{3 \Delta_{\rm C}} \right) + \frac{2 \pi^2 J^2}{\Delta_{\rm C}^2 S^2} + \frac{9 C^2 \Delta_{\rm C}^2}{32 P^2 S^2} \right] \, .
	\end{align}
	Here, $M$, $V$, $Q$, $J$, and $S$ denote the mass, thermodynamic volume, electric charge, angular momentum, and horizon entropy, respectively. The quantities $C$ and $\Delta_{\rm C}$ correspond to the average and differential conical deficits of the spacetime, respectively (see Eq.~(9) of \cite{Gregory:2019dtq}). For our purposes, it is important to note that these parameters satisfy $C\geq0$ and $\Delta_{\rm C}\geq0$. We also introduce the shorthand notation 
	\be 
	x \equiv \frac{8 P S}{\Delta_{\rm C}} \, ,
	\ee
	which was introduced in Eq.~(13) of \cite{Gregory:2019dtq} and subsequently used in Eq.~(17) therein. Using the Christodoulou--Ruffini mass formula given in Eq.~(17) of \cite{Gregory:2019dtq}, one obtains
	\be
	\frac{4 \pi M^2}{S} 
	\le \left ( \frac{3\pi M V}{2 S^2} - \frac{2C^2}{x^2} \right)^2
	-4 \left (\frac{\pi J}{S}\right)^4\, ,
	\ee
while Eqs.~(11) and (13) of \cite{Gregory:2019dtq} imply
	\be 
	\frac{3\pi M V}{2 S^2} - \frac{2C^2}{x^2} >   0 \, .
	\ee
Combining these relations and using $S=A/4$, we obtain the following identity \cite{Amo:2023bbo}:
	\be\label{C-metiden} 
	36\pi M^2V^2 - M^2A^3 -64\pi^3J^4  \ge  0 \, .
	\ee 
	
\section{Pairwise-equal four-charge AdS black hole}\label{pefcblackhole}	
	
These black holes were first constructed in \cite{Chong:2004na}, where the explicit form of the metric can be found\footnote{Here, the index $(1)$ denotes that $A_{(1)i}$ is a one-form gauge potential, while $i=1,2$ labels the two independent gauge fields.} 
	\bea\label{Pairwisemetric} 
	&\!\!\!\!\!\!\!\!\!\!&ds^2 = -\frac{\Delta_r}{W} \lp dt - \frac{a \sin^2\theta}{\Xi}d\phi \rp^2 + W\lp\frac{dr^2}{\Delta_r}+ \frac{d\theta^2}{\Delta_\theta} \rp + \frac{\Delta_\theta \sin^2\theta}{W}\lp adt - \frac{r_1 r_2+a^2}{\Xi} d\phi \rp^2,\nn\\
	&\!\!\!\!\!\!\!\!\!\!&A_{(1)1} = \frac{2r_2\sqrt{2 q_1 (m+q_1)}}{W} \lp dt - \frac{a \sin^2\theta}{\Xi} d\phi \rp,
	\nn\\
	&\!\!\!\!\!\!\!\!\!\!&A_{(1)2} = \frac{2r_1\sqrt{2 q_2 (m+q_2)}}{W} \lp dt - \frac{a \sin^2\theta}{\Xi} d\phi \rp,
	\eea
	where
	\bea\label{Pairwisepara}
	&&W = r_1 r_2 + a^2 \cos^2\theta, \quad \Delta_r = r^2 + a^2 - 2 m r + g^2 r_1 r_2(r_2 r_2 +a^2 ) , \quad \Delta_\theta = 1 - g^2 a^2 \cos^2\theta,\nn\\
	&&r_1 = r_+ + 2 q_1 \, , \quad r_2 = r_+ + 2 q_2 \, , \quad \Xi = 1 - a^2 g^2 \, , \quad g \equiv \frac{1}{\ell} \, ,
	\eea
	where $\ell$ denotes the cosmological length scale. A physically admissible solution requires the charges $Q_1$ and $Q_3$ to be real and $\Xi>0$, the latter condition being equivalent to $|a|<\ell$.
	
	The extended thermodynamic properties of these black holes were studied in \cite{Cvetic:2010jb,Cvetic:2005zi}. The relevant thermodynamic quantities are given by
	\bea\label{mca}
	&&M = \frac{m+q_1+q_2}{\Xi^2} \, , 
	\quad 
	Q_1 = Q_2 = \frac{\sqrt{q_1(q_1+m)}}{2 \Xi} \, , 
	\quad  
	Q_3 = Q_4 = \frac{\sqrt{q_2(q_2+m)}}{2 \Xi}  \, ,\\
	&&J = \frac{a (m+q_1+ q_2)}{\Xi^2}, \, \quad S=\frac{\pi\lp r_2 r_2 +a^2 \rp}{\Xi}, \, \quad T=\frac{\Delta'_{r}(r = r_+)}{4\pi\lp r_2 r_2 +a^2\rp},\nn\\
		&& \Omega=\frac{a\lp1+g r_2 r_2\rp}{r_2 r_2 +a^2} ,\, \quad \Phi_1=\Phi_2=\frac{2r_1\sqrt{q_1(q_1+m)}}{r_2 r_2 +a^2} ,\, \quad \Phi_3=\Phi_4=\frac{2r_2\sqrt{q_2(q_2+m)}}{r_2 r_2 +a^2}.\nn 
	\eea
	
Using the first law of black hole thermodynamics,
\be
dM=TdS+\Omega dJ+\sum_{i=1}^{2}\Phi_i dQ_i-VdP,
\ee
	 the thermodynamic volume can be expressed as \cite{Amo:2023bbo}
	\bea\label{pethermovol}
	V &=& \frac{2 \pi }{3 \Xi^2 r_+} \left\{2 r_+ \Xi (r_+ + q_1 + q_2)(a^2+r_1 r_2) \right.\nn\\&&\left. +  a^2 \ls 2r_+(q_1 + q_2) + r_+^2 + g^2r_1^2 r_2^2 + a^2 (1+g^2 r_1 r_2)\rs\right\}\, .
	\eea
	
\section{Super-entropic Kerr-Newman-AdS black hole}\label{SUKNblackhole}

By taking a novel ultraspinning limit of the KN-AdS solution (\ref{KN-AdSmetric}), a class of black holes known as super-entropic black holes is obtained \cite{Hennigar:2014cfa}. The limit is implemented by introducing the rescaled azimuthal coordinate $ \psi=\phi/\Xi, $ followed by taking the ultraspinning limit $ a\to \ell $. The resulting geometry is a solution of the Einstein-Maxwell equations and possesses a non-compact horizon with finite area. The corresponding metric is given by \cite{Hennigar:2014cfa}
\bea\label{KNADS2}
&&ds^2=-\frac{\Delta_r}{\rho^2}\left[dt-\ell\sin^2\!\theta d\psi\right]^2
+\frac{\rho^2}{\Delta_r} dr^2+\frac{\rho^2}{\sin^2\!\theta}d\theta^2\nonumber+\frac{\sin^4\!\theta}{\rho^2}\left[\ell dt-\left(r^2+\ell^2\right)d\psi\right]^2\,,\nn\\
&&A=-\frac{qr}{\rho^2}\left(dt-\ell\sin^2\!\theta d\psi\right)\,
\eea
where
\be
\rho^2=r^2+\ell^2\cos^2\!\theta\,,\quad \Delta_r=\left(\ell+\frac{r^2}{\ell}\right)^2-2mr+q^2.
\ee
In this form, $\psi$ is a non-compact coordinate and is compactified by imposing the identification
\be\label{psiC}
\psi \sim \psi + \mu,
\ee
where $\mu$ is a dimensionless parameter. The metric (\ref{KNADS2}) can be shown to be equivalent to the solution obtained in Refs.~\cite{Gnecchi:2013mja, Klemm:2014rda} through the coordinate transformation
\be
\tau=t,\quad p=\ell\cos\theta,\quad \sigma=-\frac{\psi}{\ell},\quad L=\frac{\mu}{\ell}.
\ee

The thermodynamic parameters of the super-entropic black hole are given by
\begin{align}\label{suentthermoqu}
	M&=\frac{\mu m}{2\pi}\,,\quad J=M\ell\,,\quad 
	\Omega = \frac{\ell}{r_+^2+\ell^2}\,,\quad T=\frac{1}{4\pi r_+}\left(3\frac{r_+^2}{\ell^2}-1-\frac{q^2}{\ell^2+r^2} \right), \nonumber \\ 
	S&= \frac{\mu}{2}\left(\ell^2+r_+^2\right),\quad \Phi=\frac{qr_+}{r_+^2+\ell^2}, \;\;\; Q=\frac{\mu q}{2\pi}\, .
\end{align}

The thermodynamics of the super-entropic black hole can be formulated in the extended phase space by promoting the cosmological constant to a thermodynamic pressure and introducing its conjugate volume. In addition, the compactification parameter $\mu$ can be incorporated as a thermodynamic variable. In the context of asymptotically Schr\"odinger geometries, the compactified null direction has been interpreted in terms of a chemical potential \cite{Herzog:2008wg}. As discussed in \cite{Klemm:2014rda}, $\psi$ is identified with a compact null coordinate on the conformal boundary. Since $\mu$ characterizes the compactification of $\psi$, it is naturally associated with a chemical-potential-like thermodynamic parameter, with $K$ denoting its conjugate quantity.

The quantities $K$ and the thermodynamic volume $V$ are determined by requiring consistency of the extended first law,
\be\label{1stlaw}
dM = TdS + VdP + \Omega dJ + \Phi dQ + Kd\mu.
\ee
This gives
\bea\label{eq}
V= \frac{r_+A}{3}
=\frac{2}{3}\mu r_+\left(r_+^2+\ell^2\right),\qquad
K=
\frac{(\ell^2-r_+^2)
	\bigl[(r_+^2+\ell^2)^2+q^2\ell^2\bigr]}
{8\pi\ell^2r_+(r_+^2+\ell^2)}.
\eea
These thermodynamic quantities satisfy the Smarr relation. Since $\mu$ is dimensionless, the term $K\mu$ does not contribute to the Smarr relation.
	
\section{Derivation of the $D$-dimensional inequalities}\label{Ddimensionalineq}

Here, the four $D$-dimensional inequalities used in the main text are derived from the intermediate refined reverse isoperimetric inequalities introduced in \cite{Amo:2023bbo}. It should be emphasized that the derivation below assumes the validity of the reverse isoperimetric inequality in arbitrary dimensions. Accordingly, the resulting relations should be regarded as conjectural higher-dimensional extensions rather than as independent proofs based solely on cosmic censorship.

{\it Even dimensions with multiple angular momenta.} In this case, the intermediate refined reverse isoperimetric inequality is given by \cite{Amo:2023bbo}
\begin{equation}
	\mathcal{R}^{D-1}
	\geq
	\left[
	1-
	\left(
	\frac{2\pi(D-2)J_{\min}^2}
	{(D-1)MV}
	\right)^2
	\right]^{-1/2}.
	\label{eq:intermediate-even-multi}
\end{equation}
Requiring the reverse isoperimetric inequality, ${\cal R}\geq 1$, to hold for rotating AdS black holes in $D$ dimensions leads to
\begin{equation}
		\frac{2\pi(D-2)J_{\min}^2}
		{(D-1)MV}<1.
	\label{eq:ineq-even-multi}
\end{equation}

{\it Even dimensions with a single angular momentum.} For the single-spin case, the corresponding intermediate refined reverse isoperimetric inequality is given by \cite{Amo:2023bbo}
\begin{equation}
	\mathcal{R}^{D-1}
	=
	\left[
	1-
	\left(
	\frac{8\pi J^2}
	{(D-1)(D-2)MV}
	\right)^2
	\right]^{-1/2}.
	\label{eq:intermediate-even-single}
\end{equation}
Consequently, requiring the validity of the RII yields
\begin{equation}
		\frac{8\pi J^2}
		{(D-1)(D-2)MV}<1.
	\label{eq:ineq-even-single}
\end{equation}

{\it Odd dimensions with multiple angular momenta.} In odd dimensions with multiple angular momenta, the intermediate refined reverse isoperimetric bound takes the form
\begin{equation}
	\mathcal{R}^{D-1}
	\geq
	\left(
	1-\frac{2\pi J_{\min}^2}{MV}
	\right)^{-\frac{D-3}{2(D-2)}}
	\left[
	1+
	\frac{2\pi(D-3)J_{\min}^2}
	{(D-1)MV}
	\right]^{-\frac{D-1}{2(D-2)}}.
	\label{eq:intermediate-odd-multi}
\end{equation}
In this case, requiring the RII to hold implies that the first factor must be real and positive,
\begin{equation}
	1-\frac{2\pi J_{\min}^2}{MV}>0,
\end{equation}
which leads to
\begin{equation}
	\frac{2\pi J_{\min}^2}{MV}<1.
	\label{eq}
\end{equation}

{\it Odd dimensions with a single angular momentum.} Finally, for the single-spin case in odd dimensions, the corresponding intermediate refined reverse isoperimetric bound is \cite{Amo:2023bbo}
\be
	\mathcal{R}^{D-1}
	\geq
	\left[
	1-
	\frac{4\pi J^2}
	{(D-1)(D-2)MV}
	\right]^{-\frac{D-3}{2(D-2)}}
	\left[
	1+
	\frac{4\pi(D-3)J^2}
	{(D-2)(D-1)^2MV}
	\right]^{-\frac{D-1}{2(D-2)}}.
	\label{eq:intermediate-odd-single}
\ee
For the RII to hold, the first factor must be real and positive. Therefore,
\begin{equation}
	1-
	\frac{4\pi J^2}
	{(D-1)(D-2)MV}>0,
\end{equation}
which gives
\begin{equation}
		\frac{4\pi J^2}
		{(D-1)(D-2)MV}<1.
	\label{eq:ineq-odd-single}
\end{equation}

\section{Singly spinning AdS thin black ring}\label{ssablackring}

The rotating AdS thin black ring in $D\geq5$, with horizon topology $S^1\times S^{D-3}$, was first constructed in \cite{Caldarelli:2008pz}, and its extended thermodynamics was subsequently investigated in \cite{Altamirano:2014tva}. The construction is based on bending a boosted thin black string, with transverse section $S^{D-3}$ and thickness $r_0$, into a circle $S^1$ of radius $R$. In the asymptotic region, the thin black ring can be described as a distributional source embedded in a global AdS background,
\be
ds^2 = -f dt^2
+ \frac{d\rho^2}{f} 
+ \rho^2 \left[d\theta^2 + \sin^2\theta\, d\Omega_{D-4}
+ \cos^2\theta\, d\psi^2\right],
\ee
where
\be
f=\left(1+\frac{\rho^2}{\ell^2} \right),
\ee
located at $\rho=R$ on the $\theta=0$ plane. The ring carries angular momentum along $\partial_\psi$, corresponding to a single rotation plane, which provides the balance against the ring tension and the gravitational potential.

The thermodynamic quantities associated with the perturbative solution for a thin black ring in AdS were derived in \cite{Caldarelli:2008pz,Altamirano:2014tva}. For completeness, they are reproduced here to verify the inequalities considered in the main text:
\begin{flalign}
	\label{brthermo}
	M &= \frac{(D-2)\omega_{D-3} r_0^{D-4}\ell}{8}
	\mathsf{R}\left(1+\mathsf{R}^2\right)^{\frac{3}{2}}, \qquad \Omega=\frac{1}{\ell}\sqrt{\frac{\lp1+\mathsf{R}^2\rp\ls1+\lp D-2\rp\mathsf{R}^2\rs}{\mathsf{R}^2\ls D-3+\lp D-2\rp\mathsf{R}^2\rs}}, &&\nn\\
	J &= \frac{\omega_{D-3} r_0^{D-4}\ell^2}{8}
	\mathsf{R}^2\sqrt{\left[1+(D-2)\mathsf{R}^2\right]
		\left[D-3+(D-2)\mathsf{R}^2\right]},  &&\nn\\
	S &= \frac{\pi\omega_{D-3}\ell r_0^{D-3}}{2}\mathsf{R}
	\sqrt{\frac{D-3+(D-2)\mathsf{R}^2}{D-4}}, \qquad
	T=\frac{(D-4)^{\frac{3}{2}}\sqrt{1+\mathsf{R}^2}}{4\pi r_0 \sqrt{ D-3+\lp D-2\rp\mathsf{R}^2}}. &&
\end{flalign}
Using these thermodynamic quantities together with the Smarr relation, the thermodynamic volume is obtained as
\be
V = \frac{\pi\omega_{D-3}r_0^{D-4}\ell^3}{D-1}
\mathsf{R}^3\sqrt{1+\mathsf{R}^2}.
\ee
Here, as in the main text, the dimensionless ratio $\mathsf{R}\equiv R/\ell$ has been introduced, where $R$ denotes the radius of the $S^1$ of the ring. The validity of the thin-ring approximation requires $r_0\ll\min{R,\ell}$. No hierarchy between $R$ and $\ell$ is assumed, and hence the analysis applies for arbitrary values of the ratio $R/\ell$.


\begin{thebibliography}{10}
	
\bibitem{Bekenstein:1973ur}
J.~D.~Bekenstein,
Phys. Rev. D \textbf{7}, 2333-2346 (1973)
doi:10.1103/PhysRevD.7.2333

\bibitem{Hawking:1975vcx}
S.~W.~Hawking,
Commun. Math. Phys. \textbf{43}, 199-220 (1975)
[erratum: Commun. Math. Phys. \textbf{46}, 206 (1976)]
doi:10.1007/BF02345020

\bibitem{Maldacena:1997re}
J.~M.~Maldacena,
Adv. Theor. Math. Phys. \textbf{2}, 231-252 (1998)
doi:10.4310/ATMP.1998.v2.n2.a1
[arXiv:hep-th/9711200 [hep-th]].

\bibitem{Kastor:2009wy}
D.~Kastor, S.~Ray and J.~Traschen,
Class. Quant. Grav. \textbf{26}, 195011 (2009)
doi:10.1088/0264-9381/26/19/195011
[arXiv:0904.2765 [hep-th]].

\bibitem{Cvetic:2010jb}
M.~Cvetic, G.~W.~Gibbons, D.~Kubiznak and C.~N.~Pope,
Phys. Rev. D \textbf{84}, 024037 (2011)
doi:10.1103/PhysRevD.84.024037
[arXiv:1012.2888 [hep-th]].

\bibitem{Kubiznak:2012wp}
D.~Kubiznak and R.~B.~Mann,
JHEP \textbf{07}, 033 (2012)
doi:10.1007/JHEP07(2012)033
[arXiv:1205.0559 [hep-th]].

\bibitem{Johnson:2014yja}
C.~V.~Johnson,
Class. Quant. Grav. \textbf{31}, 205002 (2014)
doi:10.1088/0264-9381/31/20/205002
[arXiv:1404.5982 [hep-th]].

\bibitem{Karch:2015rpa}
A.~Karch and B.~Robinson,
JHEP \textbf{12}, 073 (2015)
doi:10.1007/JHEP12(2015)073
[arXiv:1510.02472 [hep-th]].

\bibitem{AlBalushi:2020rqe}
A.~Al Balushi, R.~A.~Hennigar, H.~K.~Kunduri and R.~B.~Mann,
Phys. Rev. Lett. \textbf{126}, no.10, 101601 (2021)
doi:10.1103/PhysRevLett.126.101601
[arXiv:2008.09138 [hep-th]].

\bibitem{Visser:2021eqk}
M.~R.~Visser,
Phys. Rev. D \textbf{105}, no.10, 106014 (2022)
doi:10.1103/PhysRevD.105.106014
[arXiv:2101.04145 [hep-th]].

\bibitem{Frassino:2022zaz}
A.~M.~Frassino, J.~F.~Pedraza, A.~Svesko and M.~R.~Visser,
Phys. Rev. Lett. \textbf{130}, no.16, 161501 (2023)
doi:10.1103/PhysRevLett.130.161501
[arXiv:2212.14055 [hep-th]].

\bibitem{Ahmed:2023snm}
M.~B.~Ahmed, W.~Cong, D.~Kubiz{\v{n}}{\'a}k, R.~B.~Mann and M.~R.~Visser,
Phys. Rev. Lett. \textbf{130}, no.18, 181401 (2023)
doi:10.1103/PhysRevLett.130.181401
[arXiv:2302.08163 [hep-th]].

\bibitem{Altamirano:2014tva}
N.~Altamirano, D.~Kubiznak, R.~B.~Mann and Z.~Sherkatghanad,
Galaxies \textbf{2}, 89-159 (2014)
doi:10.3390/galaxies2010089
[arXiv:1401.2586 [hep-th]].

\bibitem{Dolan:2013ft}
B.~P.~Dolan, D.~Kastor, D.~Kubiznak, R.~B.~Mann and J.~Traschen,
Phys. Rev. D \textbf{87}, no.10, 104017 (2013)
doi:10.1103/PhysRevD.87.104017
[arXiv:1301.5926 [hep-th]].

\bibitem{Gregory:2019dtq}
R.~Gregory and A.~Scoins,
Phys. Lett. B \textbf{796}, 191-195 (2019)
doi:10.1016/j.physletb.2019.06.071
[arXiv:1904.09660 [hep-th]].

\bibitem{Feng:2017wvc}
X.~H.~Feng and H.~Lu,
Phys. Rev. D \textbf{95}, no.6, 066001 (2017)
doi:10.1103/PhysRevD.95.066001
[arXiv:1701.05204 [hep-th]].

\bibitem{Johnson:2019mdp}
C.~V.~Johnson,
Mod. Phys. Lett. A \textbf{35}, no.13, 2050098 (2020)
doi:10.1142/S0217732320500984
[arXiv:1906.00993 [hep-th]].

\bibitem{Johnson:2019wcq}
C.~V.~Johnson, V.~L.~Martin and A.~Svesko,
Phys. Rev. D \textbf{101}, no.8, 086006 (2020)
doi:10.1103/PhysRevD.101.086006
[arXiv:1911.05286 [hep-th]].

\bibitem{Klemm:2014rda}
D.~Klemm,
Phys. Rev. D \textbf{89}, no.8, 084007 (2014)
doi:10.1103/PhysRevD.89.084007
[arXiv:1401.3107 [hep-th]].

\bibitem{Hennigar:2014cfa}
R.~A.~Hennigar, D.~Kubiz{\v{n}}{\'a}k and R.~B.~Mann,
Phys. Rev. Lett. \textbf{115}, no.3, 031101 (2015)
doi:10.1103/PhysRevLett.115.031101
[arXiv:1411.4309 [hep-th]].

\bibitem{Hennigar:2015cja}
R.~A.~Hennigar, D.~Kubiz{\v{n}}{\'a}k, R.~B.~Mann and N.~Musoke,
JHEP \textbf{06}, 096 (2015)
doi:10.1007/JHEP06(2015)096
[arXiv:1504.07529 [hep-th]].

\bibitem{Gnecchi:2013mja}
A.~Gnecchi, K.~Hristov, D.~Klemm, C.~Toldo and O.~Vaughan,
JHEP \textbf{01}, 127 (2014)
doi:10.1007/JHEP01(2014)127
[arXiv:1311.1795 [hep-th]].

\bibitem{Appels:2019vow}
M.~Appels, L.~Cuspinera, R.~Gregory, P.~Krtou{\v{s}} and D.~Kubiz{\v{n}}{\'a}k,
JHEP \textbf{02}, 195 (2020)
doi:10.1007/JHEP02(2020)195
[arXiv:1911.12817 [hep-th]].

\bibitem{Frassino:2019fgr}
A.~M.~Frassino, R.~B.~Mann and J.~R.~Mureika,
JHEP \textbf{11}, 112 (2019)
doi:10.1007/JHEP11(2019)112
[arXiv:1906.07190 [gr-qc]].

\bibitem{Frassino:2015oca}
A.~M.~Frassino, R.~B.~Mann and J.~R.~Mureika,
Phys. Rev. D \textbf{92}, no.12, 124069 (2015)
doi:10.1103/PhysRevD.92.124069
[arXiv:1509.05481 [gr-qc]].

\bibitem{Frassino:2024bjg}
A.~M.~Frassino, R.~A.~Hennigar, J.~F.~Pedraza and A.~Svesko,
Phys. Rev. Lett. \textbf{133}, no.18, 181501 (2024)
doi:10.1103/PhysRevLett.133.181501
[arXiv:2406.17860 [hep-th]].

\bibitem{Mann:2025xrb}
R.~B.~Mann,
Int. J. Mod. Phys. D \textbf{34}, no.09, 2542001 (2025)
doi:10.1142/S0218271825420015
[arXiv:2508.01830 [gr-qc]].

\bibitem{Carter:1968ks}
B.~Carter,
Commun. Math. Phys. \textbf{10}, no.4, 280-310 (1968)
doi:10.1007/BF03399503

\bibitem{Caldarelli:1999xj}
M.~M.~Caldarelli, G.~Cognola and D.~Klemm,
Class. Quant. Grav. \textbf{17}, 399-420 (2000)
doi:10.1088/0264-9381/17/2/310
[arXiv:hep-th/9908022 [hep-th]].

\bibitem{Gunasekaran:2012dq}
S.~Gunasekaran, R.~B.~Mann and D.~Kubiznak,
JHEP \textbf{11}, 110 (2012)
doi:10.1007/JHEP11(2012)110
[arXiv:1208.6251 [hep-th]].

\bibitem{Hawking:1998kw}
S.~W.~Hawking, C.~J.~Hunter and M.~Taylor,
Phys. Rev. D \textbf{59}, 064005 (1999)
doi:10.1103/PhysRevD.59.064005
[arXiv:hep-th/9811056 [hep-th]].

\bibitem{Gibbons:2004uw}
G.~W.~Gibbons, H.~Lu, D.~N.~Page and C.~N.~Pope,
J. Geom. Phys. \textbf{53}, 49-73 (2005)
doi:10.1016/j.geomphys.2004.05.001
[arXiv:hep-th/0404008 [hep-th]].

\bibitem{Amo:2023bbo}
M.~Amo, A.~M.~Frassino and R.~A.~Hennigar,
Phys. Rev. Lett. \textbf{131}, no.24, 241401 (2023)
doi:10.1103/PhysRevLett.131.241401
[arXiv:2307.03011 [gr-qc]].

\bibitem{Lemos:1994xp}
J.~P.~S.~Lemos,
Phys. Lett. B \textbf{353}, 46-51 (1995)
doi:10.1016/0370-2693(95)00533-Q
[arXiv:gr-qc/9404041 [gr-qc]].

\bibitem{Lemos:1995cm}
J.~P.~S.~Lemos and V.~T.~Zanchin,
Phys. Rev. D \textbf{54}, 3840-3853 (1996)
doi:10.1103/PhysRevD.54.3840
[arXiv:hep-th/9511188 [hep-th]].

\bibitem{Dehghani:2002rr}
M.~H.~Dehghani,
Phys. Rev. D \textbf{66}, 044006 (2002)
doi:10.1103/PhysRevD.66.044006
[arXiv:hep-th/0205129 [hep-th]].

\bibitem{Bakhtiarizadeh:2025uks}
H.~R.~Bakhtiarizadeh,
Gen. Rel. Grav. \textbf{57}, no.7, 103 (2025)
doi:10.1007/s10714-025-03441-x
[arXiv:2501.09375 [gr-qc]].

\bibitem{Chong:2004na}
Z.~W.~Chong, M.~Cvetic, H.~Lu and C.~N.~Pope,
Nucl. Phys. B \textbf{717}, 246-271 (2005)
doi:10.1016/j.nuclphysb.2005.03.034
[arXiv:hep-th/0411045 [hep-th]].

\bibitem{Cvetic:2005zi}
M.~Cvetic, G.~W.~Gibbons, H.~Lu and C.~N.~Pope,
[arXiv:hep-th/0504080 [hep-th]].

\bibitem{Anabalon:2018qfv}
A.~Anabal{\'o}n, F.~Gray, R.~Gregory, D.~Kubiz{\v{n}}{\'a}k and R.~B.~Mann,
JHEP \textbf{04}, 096 (2019)
doi:10.1007/JHEP04(2019)096
[arXiv:1811.04936 [hep-th]].

\bibitem{Sen:1992ua}
A.~Sen,
Phys. Rev. Lett. \textbf{69}, 1006-1009 (1992)
doi:10.1103/PhysRevLett.69.1006
[arXiv:hep-th/9204046 [hep-th]].

\bibitem{Gibbons:1987ps}
G.~W.~Gibbons and K.~i.~Maeda,
Nucl. Phys. B \textbf{298}, 741-775 (1988)
doi:10.1016/0550-3213(88)90006-5

\bibitem{Garfinkle:1990qj}
D.~Garfinkle, G.~T.~Horowitz and A.~Strominger,
Phys. Rev. D \textbf{43}, 3140 (1991)
[erratum: Phys. Rev. D \textbf{45}, 3888 (1992)]
doi:10.1103/PhysRevD.43.3140

\bibitem{Zhang:2021wda}
M.~Zhang and J.~Jiang,
Eur. Phys. J. C \textbf{81}, no.11, 967 (2021)
doi:10.1140/epjc/s10052-021-09753-x
[arXiv:2110.09077 [gr-qc]].

\bibitem{Wu:2020cgf}
D.~Wu, P.~Wu, H.~Yu and S.~Q.~Wu,
Phys. Rev. D \textbf{102}, no.4, 044007 (2020)
doi:10.1103/PhysRevD.102.044007
[arXiv:2007.02224 [gr-qc]].

\bibitem{Sharif:2021yis}
M.~Sharif and Q.~Ama-Tul-Mughani,
Eur. Phys. J. Plus \textbf{136}, no.3, 284 (2021)
doi:10.1140/epjp/s13360-021-01270-w

\bibitem{Chrusciel:2006zs}
P.~T.~Chrusciel, D.~Maerten and P.~Tod,
JHEP \textbf{11}, 084 (2006)
doi:10.1088/1126-6708/2006/11/084
[arXiv:gr-qc/0606064 [gr-qc]].

\bibitem{Caldarelli:2008pz} M.~M.~Caldarelli, R.~Emparan and M.~J.~Rodriguez, 
JHEP \textbf{11}, 011 (2008) doi:10.1088/1126-6708/2008/11/011 [arXiv:0806.1954 [hep-th]].

\bibitem{Herzog:2008wg}
C.~P.~Herzog, M.~Rangamani and S.~F.~Ross,
JHEP \textbf{11}, 080 (2008)
doi:10.1088/1126-6708/2008/11/080
[arXiv:0807.1099 [hep-th]].

\end{thebibliography}
\end{document}